\documentclass[fleqn,usenatbib]{mnras}

\usepackage{newtxtext,newtxmath}
\usepackage[T1]{fontenc}
\usepackage{graphicx}	
\usepackage{amsmath}	
\usepackage{deluxetable}

\title[Ly$\alpha$ Halo around Reddest ERQ]{Probing the Inner Circumgalactic Medium and Quasar 
Illumination around the Reddest `Extremely Red Quasar' (ERQ)}

\author[M. W. Lau et al.]{
Marie Wingyee Lau,$^{1}$\thanks{E-mail: wingyeel@ucr.edu (MWL)}
Fred Hamann,$^{1}$
Jarred Gillette,$^{1}$
Serena Perrotta,$^{2}$
David S. N. Rupke,$^{3}$
\newauthor Dominika Wylezalek $^{4}$ 
and Nadia L. Zakamska$^{5,6}$\\
$^{1}$Department of Physics \& Astronomy, University of California, Riverside, CA 92521, USA\\
$^{2}$Center for Astrophysics and Space Sciences, University of California, San Diego, CA 92093, 
USA\\
$^{3}$Department of Physics, Rhodes College, Memphis, TN 38112, USA\\
$^{4}$Astronomical Calculation Institute, University of Heidelberg, D-69120 Heidelberg, Germany\\
$^{5}$Department of Physics \& Astronomy, Johns Hopkins University, Baltimore, MD 21218, USA\\
$^{6}$Institute for Advanced Study, Princeton, NJ 08540, USA\\}

\date{Accepted XXX. Received YYY; in original form ZZZ}

\pubyear{2022}

\begin{document}
\label{firstpage}
\pagerange{\pageref{firstpage}--\pageref{lastpage}}
\maketitle

\begin{abstract}
Dusty quasars might be in a young stage of galaxy evolution with prominent quasar feedback. A recently discovered population of luminous, extremely red quasars at $z\sim$~2--4 has extreme spectral properties related to exceptionally powerful quasar-driven outflows. We present Keck/KCWI observations of the reddest known ERQ, at $z=$\,2.3184, with extremely fast [\ion{O}{III}]~$\lambda$5007 outflow at $\sim$6000~km~s$^{-1}$. The Ly$\alpha$ halo spans $\sim$100~kpc. The halo is kinematically quiet, with velocity dispersion $\sim$300~km~s$^{-1}$ and no broadening above the dark matter circular velocity down to the spatial resolution $\sim$6~kpc from the quasar. We detect spatially-resolved \ion{He}{II}~$\lambda$1640 and \ion{C}{IV}~$\lambda$1549 emissions with kinematics similar to the Ly$\alpha$ halo and a narrow component in the [\ion{O}{III}]~$\lambda$5007. Quasar reddening acts as a coronagraph allowing views of the innermost halo. A narrow Ly$\alpha$ spike in the quasar spectrum is inner halo emission, confirming the broad \ion{C}{IV}~$\lambda$1549 in the unresolved quasar is blueshifted by $2240$~km~s$^{-1}$ relative to the halo frame. We propose the inner halo is dominated by moderate-speed outflow driven in the past and the outer halo dominated by inflow. The high central concentration of the halo and the symmetric morphology of the inner region are consistent with the ERQ being in earlier evolutionary stage than blue quasars. The \ion{He}{II}~$\lambda$1640/Ly$\alpha$ ratio of the inner halo and the asymmetry level of the overall halo are dissimilar to Type~II quasars, suggesting unique physical conditions for this ERQ that are beyond orientation differences from other quasar populations. We find no evidence of mechanical quasar feedback in the Ly$\alpha$-emitting halo.
\end{abstract}


\section{Introduction}


Quasars are phenomena of rapid accretion onto supermassive black holes in the centres of massive 
galaxies, that can regulate the evolution of their host galaxies via feedback processes. 
Popular models of galaxy evolution predict that supermassive black holes initially grow in 
obscurity, deep inside dusty galactic starbursts, until they are massive enough to power quasars 
\citep{Sanders+88,Hopkins+08a}. 
Outflows from the quasars then drive major blowouts of gas and dust, which quench star formation 
and reveal visibly luminous quasars in the galactic nuclei \citep{HopkinsElvis10,Liu+13}. 
At high redshifts, gaseous infalls from the intergalactic medium, e.g.\ cold-mode accretion, may 
be the dominant mechanism for feeding the build-up of galaxy masses, triggering starbursts, and 
fueling quasar activities \citep{Keres+09,Dekel+09,Fumagalli+14}. It is 
likely that infall and outflow occur together whenever high-redshift galaxy assembly 
via cold-mode accretion is accompanied by major starburst and quasar activity  
\citep{CostaSijackiHaehnelt14,Nelson+15,Suresh+19}.

Dust obscured, reddened quasars provide important tests of quasar/galaxy evolution models, because 
they are expected to be young objects, appearing during the brief blowout, transition phase 
between the initial dusty starbursts and later normal blue quasars. Dusty quasars are prime 
candidates for searching for signs of early stages of evolution. Such signs may include higher 
supermassive black hole accretion rates relative to Eddington, more common or more powerful 
outflows and feedback driven by the quasars, and/or more active inflows from the intergalactic 
medium. Various dusty quasar populations have been selected using a range of colour and brightness 
criteria, in e.g.\ \cite{CanalizoStockton01}, \cite{UrrutiaLacyBecker08}, \cite{Glikman+22}, 
\cite{Assef+15}, and \cite{Banerji+15}. All of these populations appear to 
require more extreme, powerful physical properties beyond what could be attributed to orientation 
effects in the unified model for active galactic nuclei \citep{Antonucci93,Netzer15}. 
Dusty quasars are unique laboratories to study quasar/galaxy evolution also by the virtue of being 
the dominant population of the first quasars. It is predicted that $>$50 per cent of quasars at 
the epoch of reionization have most of their UV radiation obscured by dust 
\citep{DiMascia+21,Ni+22}. Because cosmological dimming has a steep dependence on redshift and the 
first quasars are rare,  
dusty quasars at the highest redshifts are still missing in current systematic searches 
\citep{Connor+19,Vito+19}. 

\cite{Hamann+17} followed up the initial work of \cite{Ross+15} and refined the discovery of a 
population of `extremely red quasars' (ERQs). ERQs are selected from sources in the twelfth data 
release of the Baryon Oscillation Spectroscopic Survey \citep[BOSS,][]{Paris+17} in the Sloan 
Digital Sky Survey-III \citep[SDSS,][]{Eisenstein+11} matched to sources in the ALLWISE data 
release \citep{Cutri+13} of the Wide-field Infrared Survey Explorer 
\citep[WISE,][]{Wright+10}. More than 300 sources are selected, based on extremely red colour from 
the rest-frame UV to mid-IR of $i-W3>4.6$ AB mag. ERQs are at cosmic noon redshifts 
$z\sim$\;2\textendash4, and have high bolometric luminosities 
$L_{\rm bol}\gtrsim10^{47}$\,erg\,s$^{-1}$. 

The BOSS rest-frame UV spectra reveal that ERQs have exotic spectral properties unlike any other 
known quasar populations. They have a high incidence of blueshifted broad absorption lines and 
large emission-line blueshifts. They have a high incidence of unusually strong broad emission 
lines with \ion{C}{IV}\,$\lambda$1549 rest equivalent width often exceeding 100\,\AA, and the 
broad emission lines tend to have peculiar profiles that are wingless with high kurtosis 
\citep{MonadiBird22}. They have a high incidence of unusual broad emission line flux ratios such 
as high \ion{N}{V}\,$\lambda$1240/\ion{C}{IV}\,$\lambda$1549 and high  
\ion{N}{V}\,$\lambda$1240/\ion{H}{I}\,Ly$\alpha$. These features emphasize small-scale outflow 
phenomena controlled by accretion physics, on scales of tens of pc 
\citep{Zhang+17,Alexandroff+18}. Rest-frame optical spectra in \cite{Zakamska+16} and 
\cite{Perrotta+19} reveal that ERQs have the broadest and most blueshifted 
[\ion{O}{III}]\,$\lambda$5007 emission lines ever reported, reaching $>$6000\,km\,s$^{-1}$ and are 
strongly correlated with the red colours. The [\ion{O}{III}] lines are significant because they 
trace gas on galactic scales, being low-density forbidden transitions \citep{Hamann+11}. These 
[\ion{O}{III}]-emitting outflows carry enough kinetic energy to drive important feedback in the 
host galaxies. The extreme [\ion{O}{III}]\,$\lambda$5007 and broad emission line properties in 
ERQs appear to require more extreme physical properties than other quasars that are beyond 
orientation differences. 

\cite{Hamann+17} also note that narrow Ly$\alpha$ emission `spikes' are often seen in the BOSS 
spectra. The origin of these unusual spikes are speculated to be gas on larger scales than the 
broad-line region. Their high incidence in ERQ spectra is possibly caused by extinction of the 
direct central quasar light. The spikes are then important redshift indicators, especially in ERQs 
where broad emission line blueshifts appear common and estimates of the systemic redshifts from 
the broad lines are unreliable.

The known prodigious outflows support the hypothesis that ERQs are candidate young quasars, tied 
to an early, powerful stage of quasar-galaxy evolution, as have been inferred for the hot 
dust-obscured galaxies \citep[hot DOGs,][]{Noboriguchi+19,Finnerty+20}. Observations of ERQ 
host galaxies and their extended environments are needed to test this hypothesis. The host 
galaxies of ERQs are explored in \cite{Zakamska+19} using direct {\it Hubble Space Telescope} 
images with a goal of identifying signs of early stages of evolution. Somewhat surprisingly, they 
found no evidence of enhanced major merger activities relative to luminosity- and redshift-matched 
blue quasars. However, disturbances in the host galaxies may be hidden by the inherent difficulty 
in subtracting the quasar point-spread function in two-dimensional images especially at high 
redshifts. 
\cite{Vayner+21} used adaptive optics-assisted integral field spectroscopy to confirm that the 
[{O}{III}]-emitting outflows happen on $\sim$1\,kpc scales and carry enough momenta to clear the 
nuclear regions of their host galaxies. To explore whether the ERQ-driven outflows have detectable 
impact on the larger, circumgalactic scales, one may employ wide-field integral field 
spectrographs, such as the Keck Cosmic Web Imager \citep[KCWI,][]{Morrissey+18} on the Keck II 
telescope. The circumgalactic medium is loosely defined as the gaseous halo extending out to 
$\sim$300\,kpc from the central galaxy \citep{QPQ8,TumlinsonPeeplesWerk17}. 
It is the site of interplay between outflows from and accretion onto the galaxy it surrounds, 
and has significant impact on the evolution of the galaxy. 

Massive gas reservoirs around high-redshift quasars have been detected via fluorescent Ly$\alpha$ 
emission. Around mostly radio-quiet, luminous blue quasars, emission-line regions of linear 
extents $\sim$100\,kpc are frequently detected around statistical samples of them, using 
wide-field integral field spectrographs on 8m class telescopes 
\citep{Borisova+16a,ArrigoniBattaia+19,Cai+19}, 
They are often interpreted as tracing filamentary, inflowing gas of cold-mode accretion. 
For radio-quiet obscured quasars, extended line emissions have been detected around sources 
selected using various methods, using mixed observing techniques on telescopes of various 
apertures to various exposure depths. 
\citep{Bridge+13,PrescottMartinDey15,Prescott+15b,Cai+17,denBrok+20,HerenzHayesScarlata20,OSullivan+20,Li+22,Sanderson+21}. 
ERQs selected from the BOSS quasar catalogue are a well-defined and uniform sample. They are thus 
ideal targets to advance Ly$\alpha$ halo studies beyond random surveys 
\citep[e.g.,][]{Borisova+16a,ArrigoniBattaia+19,Cai+19} 
or exotic single objects \citep[e.g.,][]{Cantalupo+14,Ginolfi+18,Li+19} to tests of quasar/galaxy 
evolution that are both statistical and specific. 

Our team is currently building a KCWI sample of ERQs. 
In this paper we report results on the reddest of all known ERQs, SDSS J000610.67+121501.2, 
hereafter J0006+1215. It has a rest-frame UV to mid-IR colour of $i-W3=8.01$ AB mag, an emission 
redshift of $z_{\rm em}=2.3184$, and is radio quiet. The first goal relates to the 
circumgalactic medium and quasar illumination pattern. We will test whether the extended line 
emissions have unusual properties that may relate to the quasar obscuration, and/or a younger, 
more active, blowout phase of evolution. We will search for differences compared to the more 
extensively studied luminous blue quasars at similar redshifts, which are expected to be more 
evolved. The second goal relates to using the Ly$\alpha$ halo to estimate the systemic redshift. 
We will test the assertion that the Ly$\alpha$ spike in the BOSS spectrum is halo emission at the 
systemic redshift, that can be used to study the apparently very large blueshifts of the broad 
emission lines. 

This paper is organized as follows. Section 2 describes the selection of J0006+1215, its 
observation, data reduction, and post-processing. Section 3 describes the spectral properties of 
the J0006+1215 quasar, extraction of the extended line emissions, and analysis results of size, 
morphology, surface brightness, and kinematics of the extended emissions. Section 4 discusses the 
implications for quasar feedback, circumgalactic medium, and quasar studies. Section 5 concludes 
the paper. Throughout this paper we adopt a $\Lambda$ cold dark matter cosmology with 
$H_0=69.6$\,km\,s$^{-1}$\,Mpc$^{-1}$, $\Omega_{\rm M}=0.286$, and $\Omega_{\rm \Lambda}=0.714$, as 
adopted by the online cosmology calculator developed by \cite{Wright06} at the time of this 
writing. We use photometric magnitudes in the AB system. We report vacuum wavelengths in the 
heliocentric frame. When referring to broad, blended doublet emission of \ion{C}{IV}, we use the
 notation \ion{C}{IV}\,$\lambda$1549, and when referring to narrow, resolved doublet emission of 
\ion{C}{IV}, we use the notation \ion{C}{IV}\,$\lambda\lambda$1548,1550. 

\section{Target Selection, Observation, and Data Reduction}

We provide details on the selection of J0006+1215, its KCWI observation, data reduction, and 
post-processing including modelling the quasar point-spread function. 

\subsection{J0006+1215 Selection and Basic Properties}

\begin{figure}
\includegraphics[width=\columnwidth]{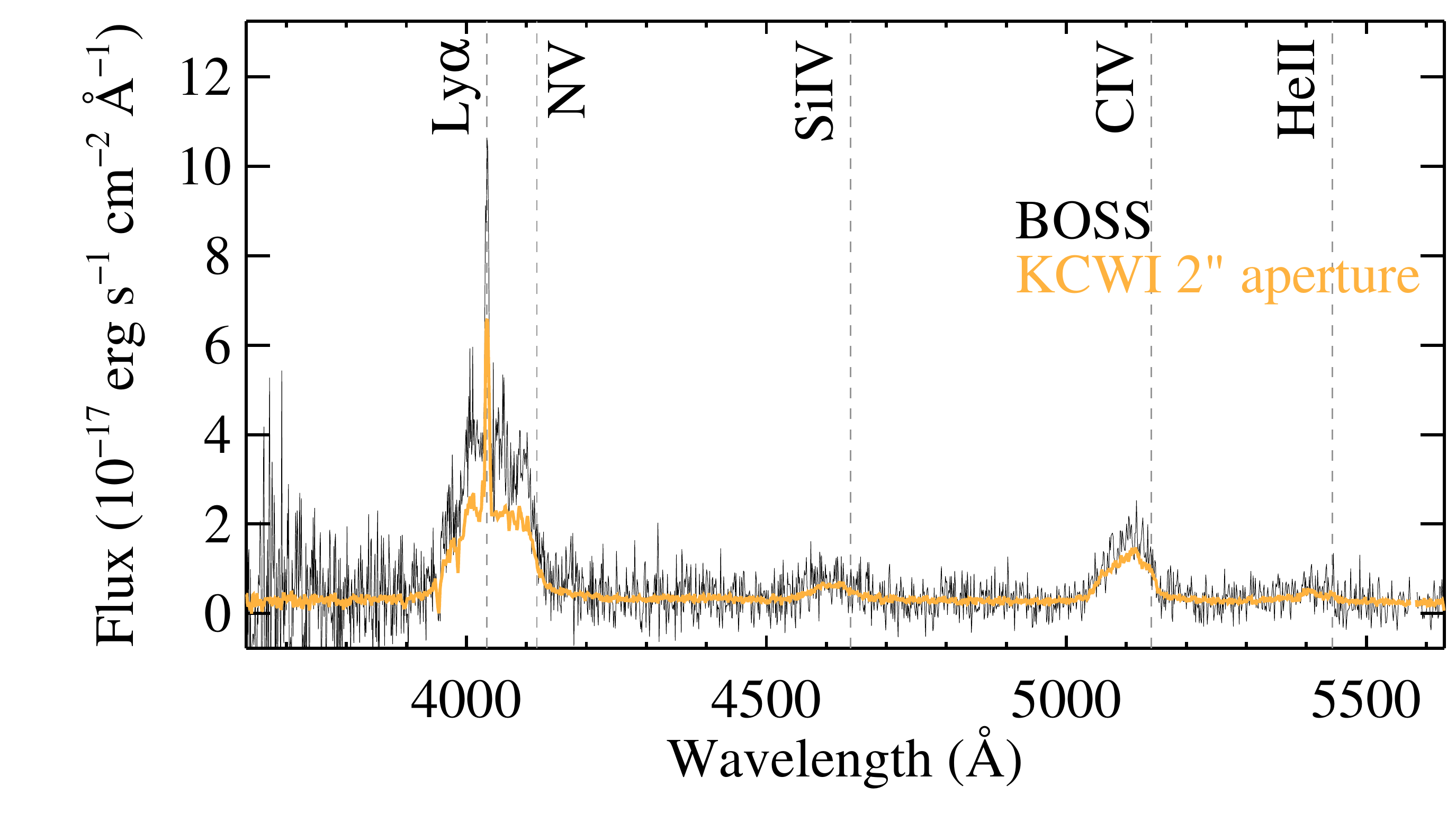}
\caption{The BOSS spectrum of the quasar taken in 2011, compared to the KCWI spectrum of the 
quasar taken in 2019 and extracted with a matching aperture.
}
\label{fig:KCWIBOSS}
\end{figure}

For this first wide-field integral-field study of an ERQ, we choose the reddest among all known 
ERQs, with $i-W3=8.01$, which also has among the most blueshifted and broadest [\ion{O}{III}] 
lines among ERQs. As measured by \cite{Perrotta+19} in a one-dimensional spectrum, the velocity 
shift that encompasses 98 per cent of its [\ion{O}{III}]\,$\lambda$5007 emission-line flux 
integrating from the red side $v_{98,{\rm [OIII]}}=-6092$\,km\,s$^{-1}$, and the velocity width 
encompassing 90 per cent of the [\ion{O}{III}] line power 
$w_{90,{\rm [OIII]}}=6206$\,km\,s$^{-1}$. For reference, the most extreme $v_{98,{\rm [OIII]}}$ 
and $w_{90,{\rm [OIII]}}$ measured in ERQs are $\sim$7000\,km\,s$^{-1}$. We note that a narrow 
emission component of Ly$\alpha$ is present in the BOSS spectrum, as shown in 
Fig.~\ref{fig:KCWIBOSS}, and this feature is extensively compared with halo properties in later 
sections. 

The absolute {\it i} magnitude {\it K}-corrected to $z\sim2$ is a commonly used measure of 
the rest-frame UV luminosity of quasars. We follow the formulae for {\it K}-corrections in 
\cite{Ross+13} and calculate that J0006+1215 has ${\rm M}_i(z=2)=-24.16$. 
\cite{Hamann+17} estimate that ERQs are typically suppressed by three magnitudes in the rest-frame 
UV compared to the spectral energy distributions of normal blue quasars. J0006+1215 is suppressed 
by about six magnitudes in the rest-frame UV, and we estimate its intrinsic bolometric luminosity 
from mid-IR photometry assuming a standard Type~I spectral energy distribution. We apply a 
bolometric correction factor of 8 to the rest-frame 5\,$\umu$m luminosity extrapolated from the 
{\it W}3 band photometry, with details in \cite{Perrotta+19}. J0006+1215 has an intrinsic 
bolometric luminosity of $L_{\rm bol}=10^{47.9}$\,erg\,s$^{-1}$, and is near the high end of the 
ERQ population. 
The radio luminosity of J0006+1215 is in the quiet regime, and is unresolved on $\sim$10\,kpc 
scale \citep{Hwang+18}. The X-ray spectrum of J0006+1215 indicates a Compton-thin absorbing 
column \citep{Goulding+18}, which is discussed in Section 4.5 regarding quasar illumination. 

\subsection{Observing}

We observed J0006+1215 with Keck/KCWI on 2019 October 2. The seeing was variable with 
full-width-at-half-maximum (FWHM) between (0.8\textendash1.4)\,arcsec, which correspond to 
(7\textendash12)\,kpc at the quasar's redshift. The observing conditions were photometric. 
We configured KCWI with the BL grating and the medium slicer, which yields a field of view of 
15.7\,arcsec $\times$ 18.9\,arcsec, with slices along the long direction. 
At the emission redshift of the quasar, this field of view corresponds to a 
physical area of 131\,kpc\,$\times$\,158\,kpc. 
With 24 slicers, the instrument configuration provides a spatial sampling of 0.68\,arcsec 
across the slices and is seeing limited along the slicers. Consecutive exposures were dithered 
0.35\,arcsec along slices to sub-sample the long spatial dimension of the output spaxels which 
have sizes of 0.68\,arcsec $\times$ 0.29\,arcsec. The spectral resolution is $R=1800$. The full 
spectral range is (3500\textendash5625)\,\AA, and covers Ly$\alpha$, \ion{C}{IV}\,$\lambda$1549, 
and \ion{He}{II}\,$\lambda$1640 of the quasar in observer's frame. The total exposure time was 70 
min. We obtained three exposures of 20 minutes, and one exposure of 10 minutes which was read 
out early due to fog. We observed an unsaturated standard star GD50 at the end of the night for 
flux calibration. 

\subsection{Reducing the Data}

We reduce the data of the individual science exposures with the standard KCWI Data Extraction and 
Reduction Pipeline (KDERP, \url{https://github.com/Keck-DataReductionPipelines/KcwiDRP}) written 
in the Interactive Data Language (IDL), whose final outputs are data cubes of two spatial 
dimensions and one spectral dimension. We then use the IDL library IFSRED \citep{Rupke14a} to 
resample and mosaic the data cubes. The output data cubes from KDERP have rectangular spaxels, and 
they are resampled onto a 0.29\,arcsec $\times$ 0.29\,arcsec spaxel grid. Spaxel coordinates of 
the centroid of the quasar is calculated for each resampled data cube. The individual data cubes 
are then aligned using the position of the quasar centroid and median-combined. There are no 
prominent skyline residuals near the lines we investigate. 

We then verify the reliability of the flux calibration. The bandpass of the KCWI data 
approximately matches the SDSS photometric {\it g} band, hence the SDSS {\it g} magnitude of the 
quasar can be used as a baseline for comparison. We extract a source spectrum from a circular 
aperture of 3\,arcsec in diameter that is centred on the quasar in the KCWI data. This 
aperture is two times the seeing disc FWHM and is sufficiently large to encompass most of the 
source emission. We then filter the KCWI source flux spectrum with the SDSS {\it g}-band response 
function and integrated over the entire wavelength range. The integrated flux is converted to 
magnitude and is found to be $g_{\rm KCWI}=22.26$, and is 8 per cent fainter compared to the 
magnitude reported in SDSS $g_{\rm SDSS}=22.16$. 
Given quasar variability, 
We verify the flux calibration using the standard star as well. For the standard star, the 
{\it g}-band flux integrated from the KCWI spectrum is merely 2 per cent smaller than the SDSS 
{\it g}-band flux. We hence only report uncertainties on the flux measurements as output from 
KDERP and assume there are no additional unaccounted systematic uncertainties in flux calibration. 
The final data cube results in a 1$\sigma$ surface brightness limit of 
$2.2\times10^{-19}$\,erg\,s$^{-1}$\,cm$^{-2}$\,arcsec$^{-2}$ in a 1-arcsec aperture in a 
single 1\,\AA\ channel at the Ly$\alpha$ wavelength at the quasar's redshift. The root-mean-square 
of surface brightness in a 1-arcsec aperture is 
$2.0\times10^{-19}$\,erg\,s$^{-1}$\,cm$^{-2}$\,arcsec$^{-2}$ for a channel in rest-frame 
wavelengths between 1255\,\AA\ and 1275\,\AA, a range with no prominent emission lines from the 
quasar. 

\subsection{Modelling the Point Spread Function and Other Post-Processing}

\begin{figure}
\includegraphics[width=1.1\columnwidth]{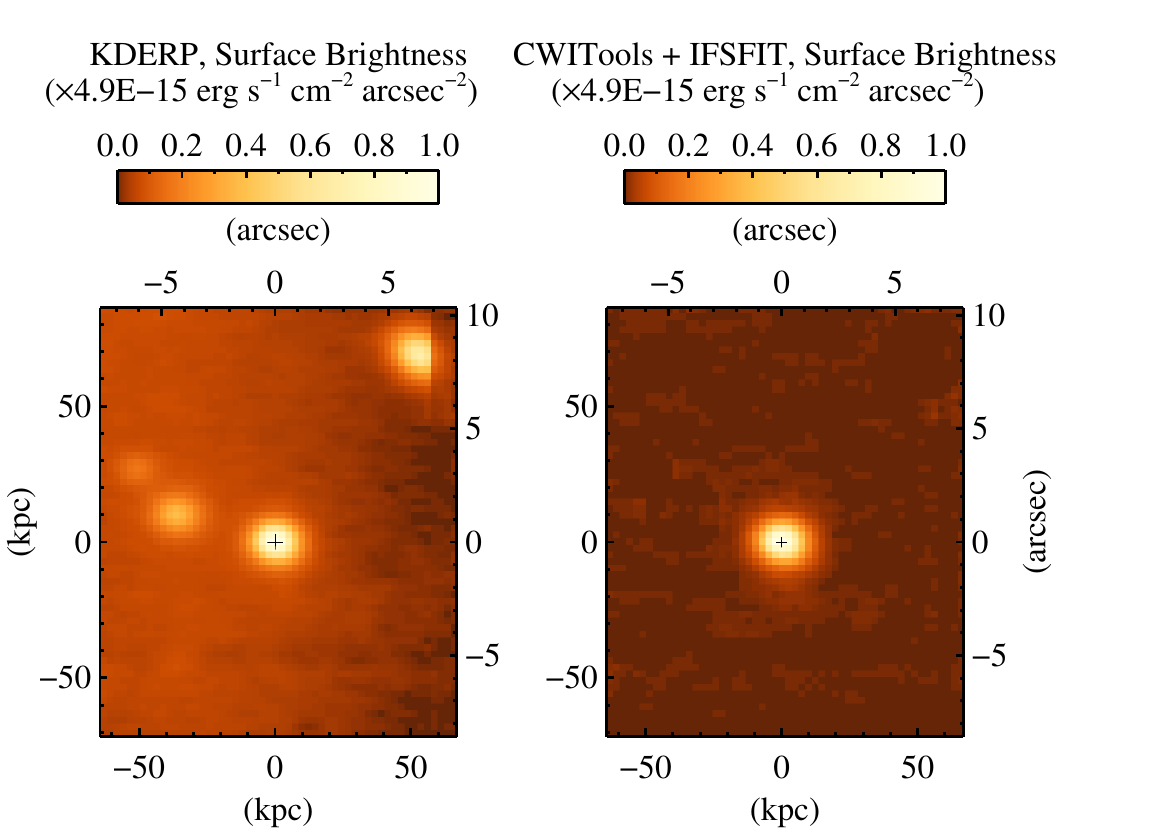}
\caption{The left panel shows a white-light image obtained from the reduced, mosaicked, spatially 
resampled data cube output by KDERP and IFSRED. The right panel shows a white-light image obtained 
from the further processed data cube that have residual background light and foreground sources 
subtracted using CWITools and IFSFIT. The cross symbol marks the peak of the quasar. Comparing the 
two images, residual background coming from imperfect calibrations is reduced and foreground 
objects are removed.}
\label{fig:whitelights}
\end{figure}

We further process the data cube using the Python package CWITools \citep{OSullivanChen20} for 
subtracting instrumental scattered light and foreground sources. We mask all continuum sources and 
median-filter the data cube as an estimate of instrumental scattered light and subtract it. On the 
basis that the point-spread function does not vary strongly as a function of wavelength, as 
empirically verified, we fit empirical point-spread functions or Moffat functions to foreground 
continuum sources. We then subtract these fits to foreground sources. We then process the spectrum 
in each spaxel of the data cube using the IDL library IFSFIT \citep{RupkeTo21}. We generate a 
spatially-unresolved quasar spectral template of continuum plus broad emission line components 
that is described in detail in Section 2.5. For each spectrum we mask emission-line regions 
of the quasar, fit the line-free regions with low-order polynomials plus a flux re-scaled 
unresolved quasar template, and then fit the narrow emission lines with Gaussian functions. The 
spatial distribution of the flux re-scaling factors of the unresolved quasar template forms a 
model for the point-spread function. The low-order polynomials account for imperfections in 
calibrations, including instrumental scattered light subtraction, sky subtraction, foreground 
source subtractions, and modelling the quasar point-spread function. To demonstrate that 
processing with CWITools and IFSFIT reduces calibration imperfections and removes foreground 
objects, in Fig.~\ref{fig:whitelights} we show spectrally integrated, white-light images obtained 
from collapsing the data cube before and after subtracting a median filter to the background 
field, foreground sources, and low-order polynomials in wavelength space. We fit a circular Moffat 
function to the empirical point-spread function and find a FWHM of 1.4\,arcsec. The size of this 
point-spread function is slightly larger than the atmospheric seeing disc, which may be attributed 
to the mosaicking procedure and variable seeing. We correct all measurements of surface 
brightnesses, spatially-integrated fluxes, and luminosities for a Galactic extinction of 
$A_V=0.214$ \citep{SchlaflyFinkbeiner11} toward the quasar location. This corresponds to a 33 per 
cent upward correction at Ly$\alpha$, a 24 per cent upward correction at 
\ion{C}{IV}\,$\lambda$1549, and a 22 per cent upward correction at \ion{He}{II}\,$\lambda$1640. 

\subsection{Isolating Narrow Emissions in the Central 1-arcsec}

\begin{figure}
\includegraphics[width=\columnwidth]{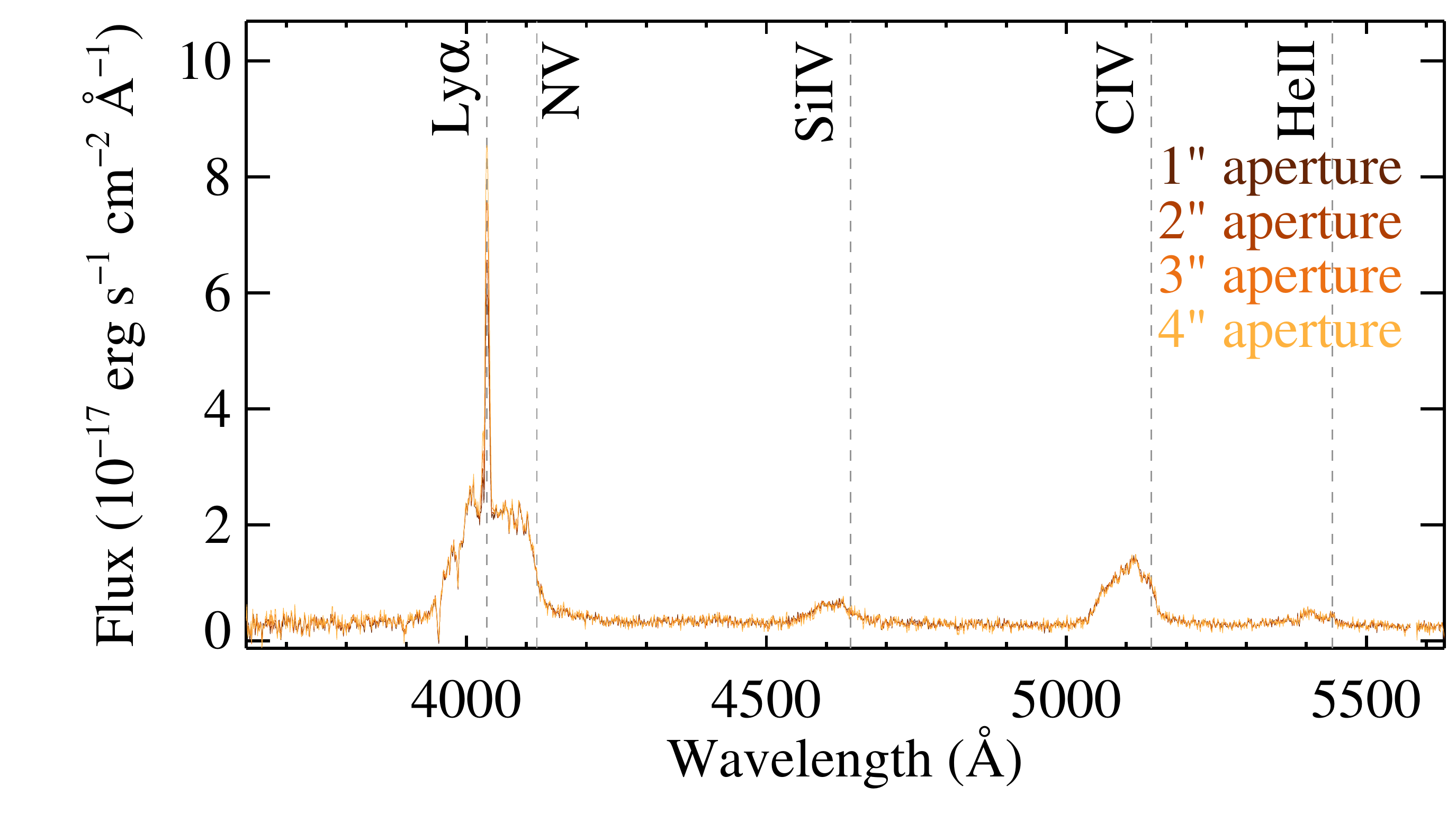}
\includegraphics[width=\columnwidth]{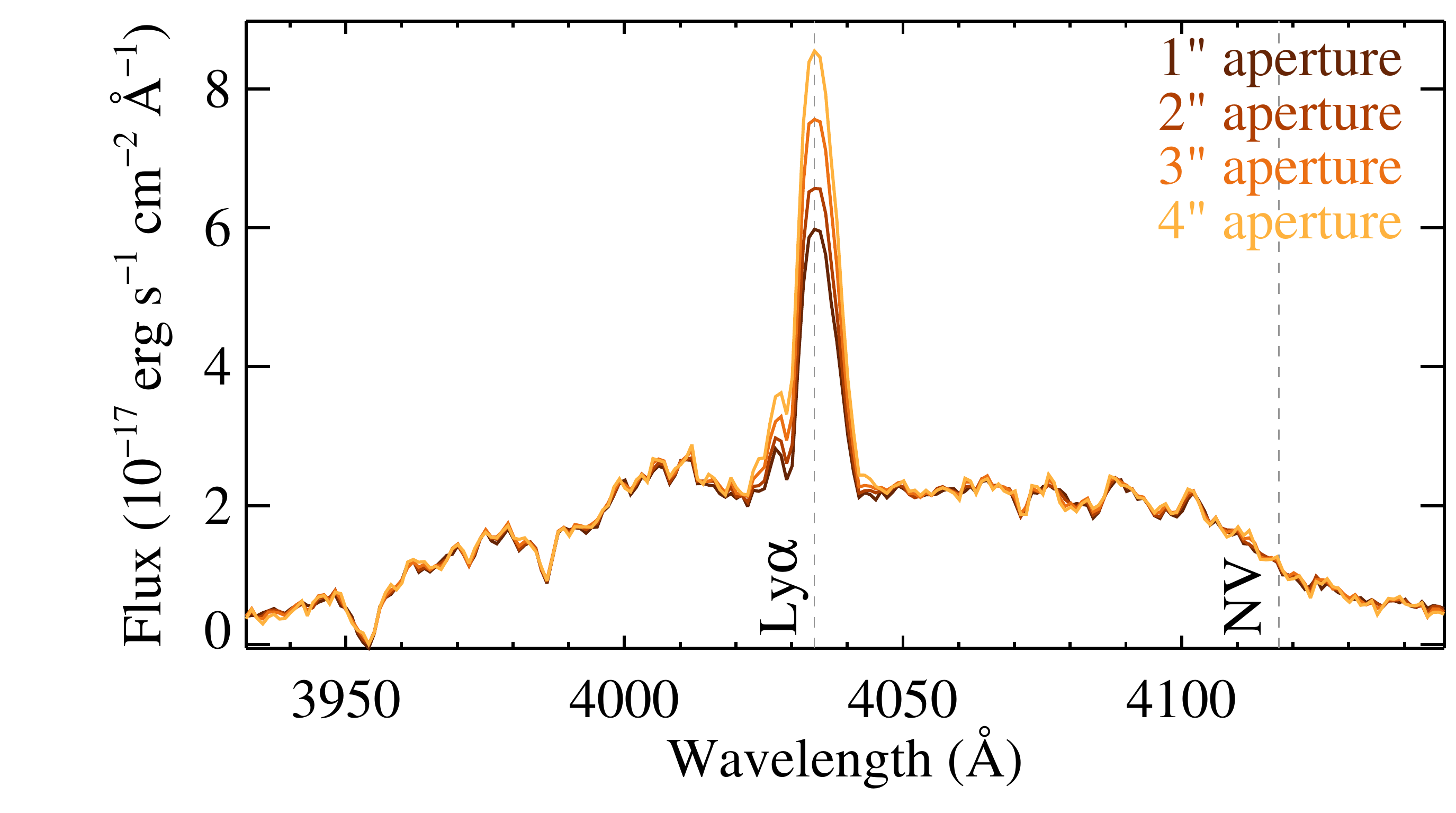}
\includegraphics[width=\columnwidth]{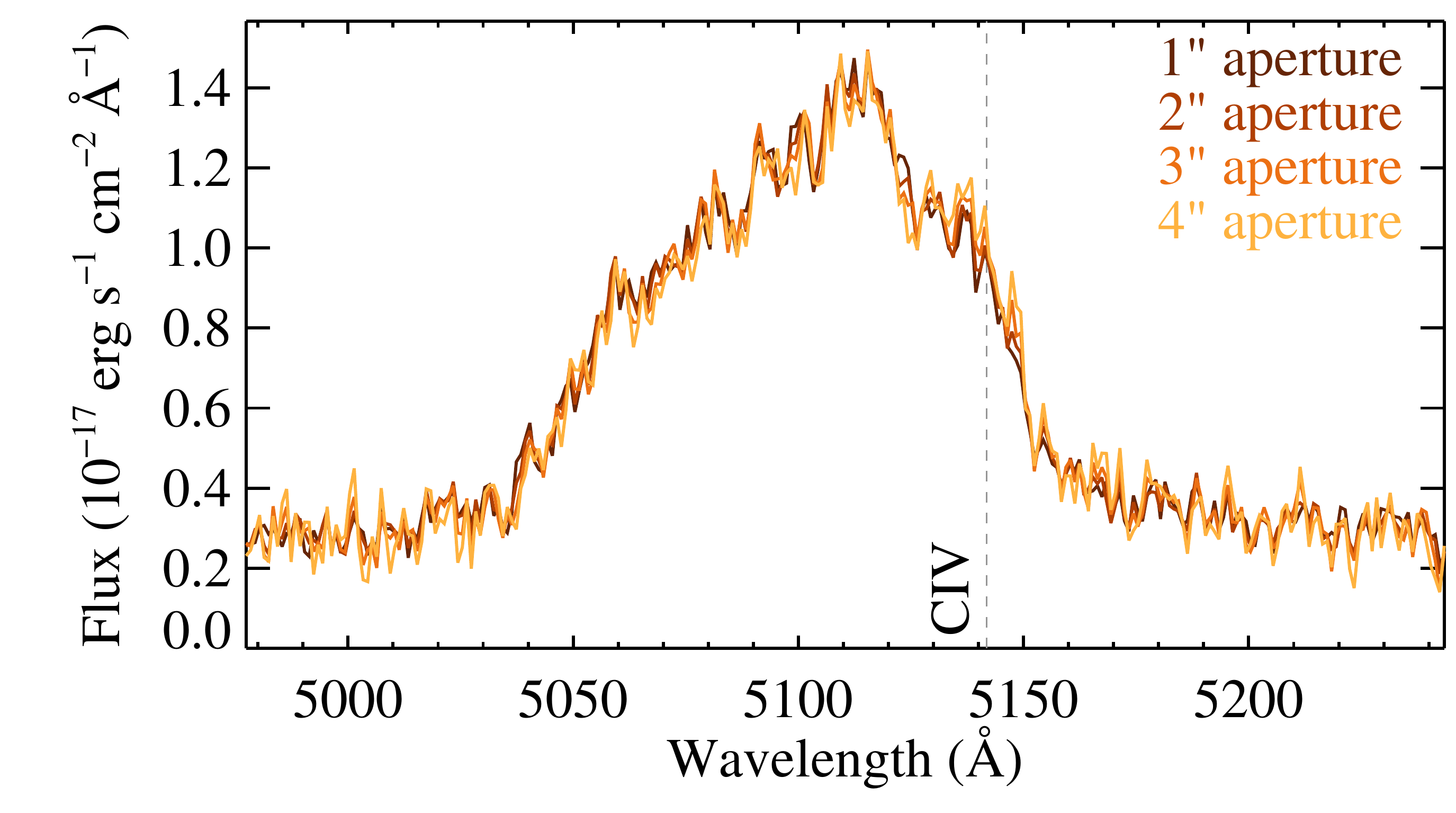}
\caption{The panels show spectra extracted from circular apertures of 1\,arcsec, 2\,arcsec, 
3\,arcsec, and 4\,arcsec in diameter and centred on the quasar, all scaled to the continuum level 
of the 2-arcsec spectrum. In the top panel the full spectra are shown. The prominent spike is 
Ly$\alpha$ emission, which defines the systemic frame. Locations of 
\ion{N}{V}\,$\lambda$1240, \ion{Si}{IV}\,$\lambda$1398, \ion{C}{IV}\,$\lambda$1549, and 
\ion{He}{II}\,$\lambda$1640 in this frame are marked with vertical dashed lines. With increasing 
aperture sizes, the flux level of the broad emission lines relative to the continuum stays 
unchanged, indicating the broad-line region is spatially unresolved. The middle panel shows a 
zoom-in of the Ly$\alpha$-\ion{N}{V} region. The flux level of the narrow Ly$\alpha$ increases 
with aperture size, indicating its emission region is spatially extended. The bottom panel shows a 
zoom-in of the \ion{C}{IV}\,$\lambda\lambda$1548,1550 region. Extended narrow line emissions are 
not very apparent in \ion{C}{IV} and \ion{He}{II}\,$\lambda$1640 and their presence is found by 
modelling. 
}
\label{fig:corespecs}
\end{figure}

In rest-frame UV spectra of blue unobscured quasars, the narrow emission lines that originate far 
from the quasar broad-line regions are much weaker than the broad emission lines and smoothly blend 
with them. Generally the spatially-unresolved quasar spectral template is generated by simply 
summing a circular aperture matching the size of the seeing disc and centred on the quasar, as the 
narrow and broad emission lines cannot be easily separated. For this study of an ERQ,  
the broad-line region is partially obscured so that narrow lines can be identified as 
kinematically distinct components. To make the unresolved quasar template we subtract the narrow 
emission lines. Fig.~\ref{fig:corespecs} displays spectra of summing 1-arcsec, 2-arcsec, 3-arcsec, 
and 4-arcsec circular apertures centred on the quasar in the data cube, all scaled to match the 
continuum level of the 2-arcsec aperture spectrum. The Earth atmosphere's glow at 
[\ion{O}{I}]\,$\lambda$5577 is omitted from all spectra. With increasing aperture sizes, the flux 
level of the broad emission in Ly$\alpha$, \ion{N}{V}\,$\lambda$1240, \ion{Si}{IV}\,$\lambda$1398, 
and \ion{C}{IV}\,$\lambda$1549 relative to the continuum remains unchanged. This confirms that the 
broad-line region is spatially unresolved, and should be part of the quasar template. On 
the contrary, the flux level of the prominent Ly$\alpha$ emission spike increases with aperture 
size. This indicates the narrow Ly$\alpha$-emitting region is not part of the spatially unresolved 
point-spread function. To make the quasar template, we clip the Ly$\alpha$ spike, and fit and 
subtract any narrow emission in \ion{C}{IV}\,$\lambda\lambda$1548,1550 and 
\ion{He}{II}\,$\lambda$1640 from a spectrum extracted from a circular aperture of 1\,arcsec 
diameter. The aperture size is determined as a compromise between including enough of the quasar's 
emission for good signal-to-noise (S/N) while also minimizing spatially extended emission. 

\begin{figure}
\includegraphics[width=\columnwidth]{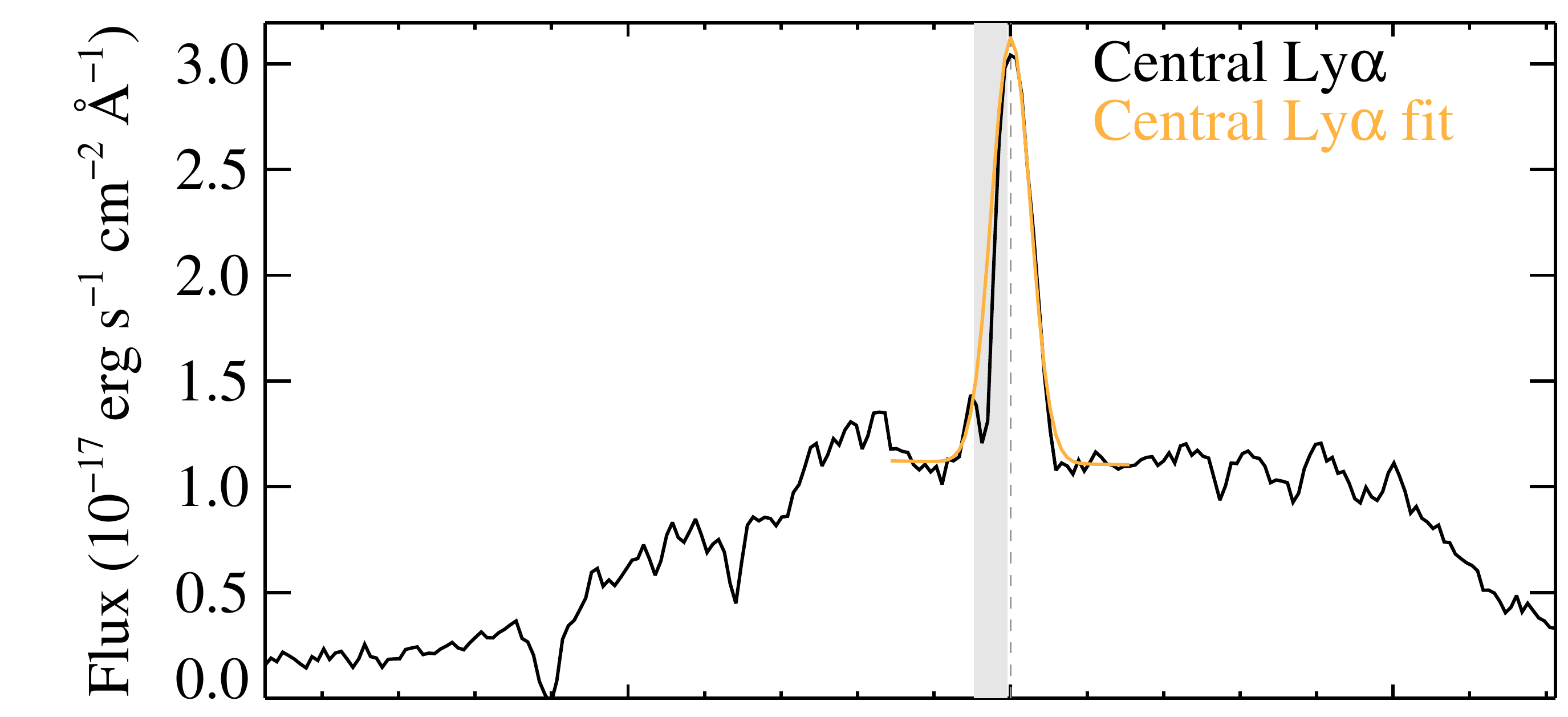}
\includegraphics[width=\columnwidth]{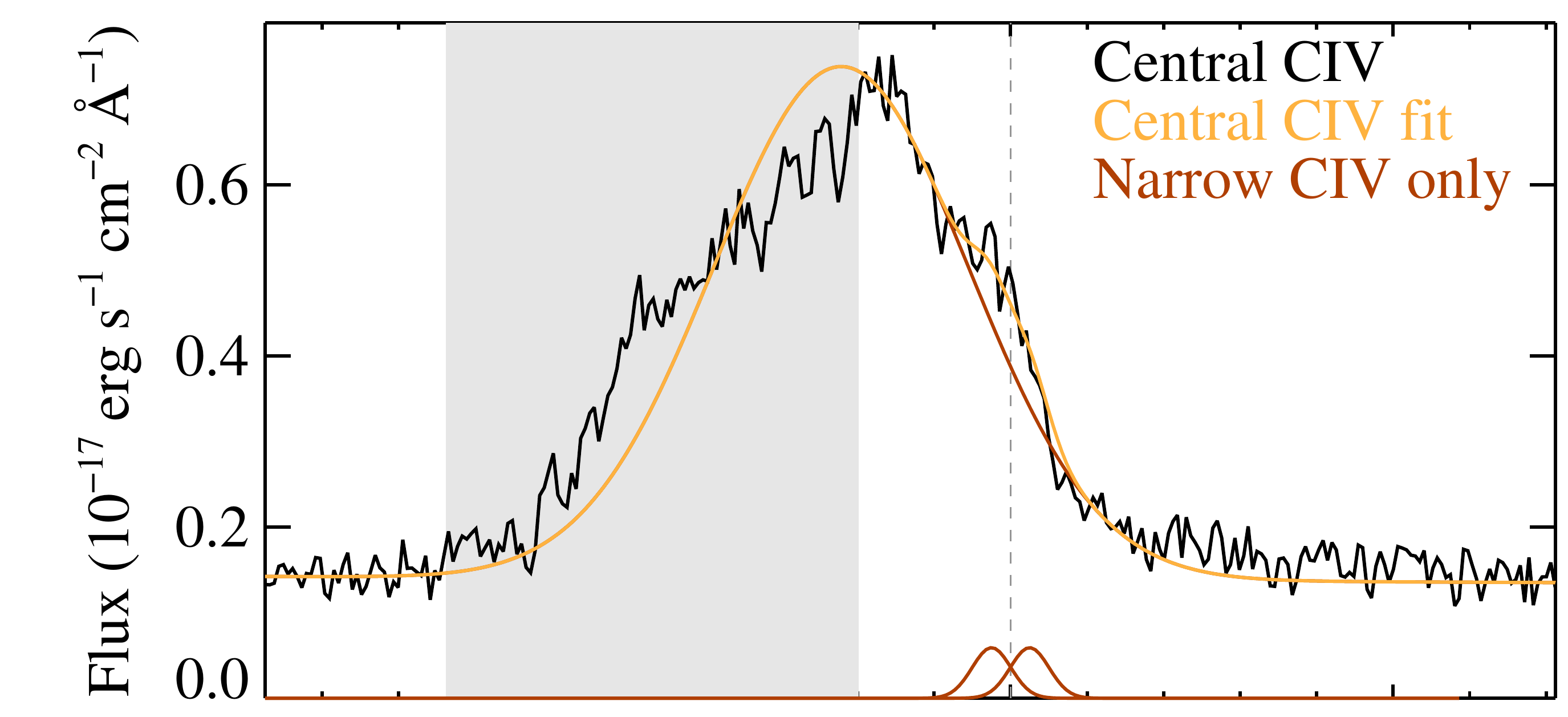}
\includegraphics[width=\columnwidth]{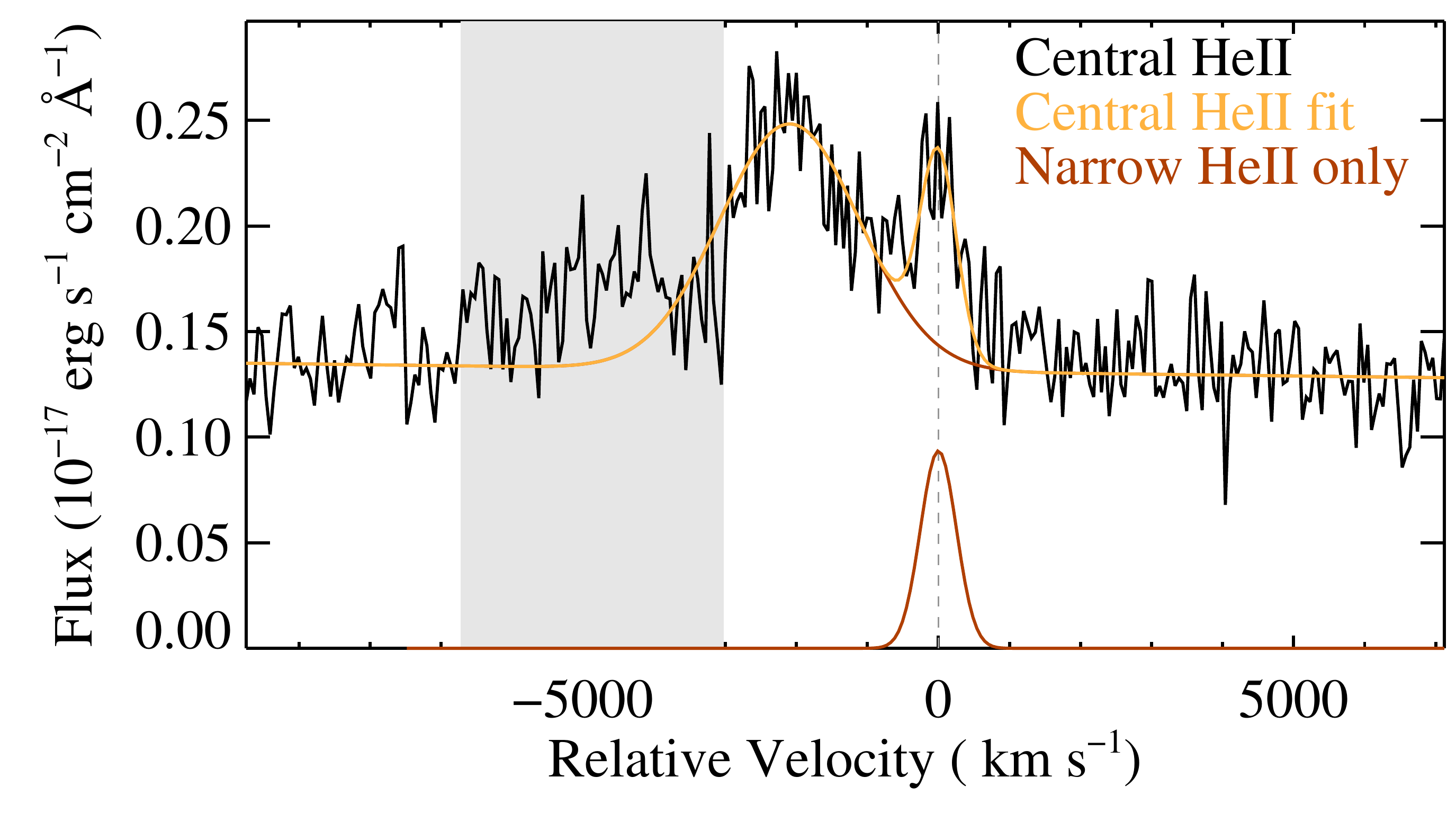}
\caption{Fitting for the narrow component of emission lines in the central 1-arcsec aperture. The 
black curves are the data, the orange curves are the fits, and the grey shades are regions masked
during fitting. Grey dashed lines mark locations of zero velocities. Top panel shows the 
Ly$\alpha$ region. The fit is the sum of a Gaussian plus a linear function. The absorption on the 
blue side of the line profile is masked. The peak of this fit is adopted as the systemic velocity. 
Middle panel shows the \ion{C}{IV}\,$\lambda$1549 region. The fit is the sum of a broad Gaussian, 
two narrow Gaussians of fixed separation, plus a continuum. The asymmetric blue side of the line 
is masked, and the kinematics of the narrow component of the 
\ion{C}{IV}\,$\lambda\lambda$1548,1550 doublet are fixed at narrow Ly$\alpha$'s. The brown curve 
separately plots the sum of the broad Gaussian plus the continuum and the sum of the two narrow 
Gaussians. Bottom panel shows the \ion{He}{II}\,$\lambda$1640 region. The fit is the sum of a 
broad Gaussian, a narrow Gaussian, plus a continuum. The blue side of the line is masked, and the 
narrow component's kinematics are fixed at narrow Ly$\alpha$'s. The brown curve separately plots 
the sum of the broad Gaussian plus the continuum and the narrow Gaussian.}
\label{fig:fitcorespec}
\end{figure}

We construct the unresolved quasar spectral template by extracting a one-dimensional spectrum from 
a 1-arcsec aperture centred on the quasar and subtracting narrow line emissions. 
Fig.~\ref{fig:fitcorespec} shows zoom-ins of the Ly$\alpha$ region, the 
\ion{C}{IV}\,$\lambda$1549 region, and the \ion{He}{II}\,$\lambda$1640 region of this spectrum. 
Overplotted on the data is our fits. We use the Ly$\alpha$ spike to determine the systemic 
redshift of the quasar host galaxy and we describe our fitting of the spike below. 
As shown in Fig.~\ref{fig:corespecs}, 
the broad Ly$\alpha$-\ion{N}{V}\,$\lambda$1240 emission complex underlying the spike in a local 
wavelength range appears linear in shape. We fit a single Gaussian component to the narrow 
Ly$\alpha$ emission plus a linear function to the part of the broad 
Ly$\alpha$-\ion{N}{V}\,$\lambda$1240 complex underlying it. 
The blue side of the narrow Ly$\alpha$ line profile has an absorption feature, and is masked in 
the fitting process. The evidence for the blue side absorption is verified in the analysis of the 
Ly$\alpha$ halo spectrum which is discussed in Section 3.4. We find $z_{\rm sys,Ly\alpha}=2.3184$ 
from the Gaussian component of the fit, and we adopt this systemic redshift for all relative 
velocity measurements. We do not attempt to fit the full Ly$\alpha$-\ion{N}{V}\,$\lambda$1240 
complex. 

Fig.~\ref{fig:fitcorespec} shows that at the systemic velocity defined by the narrow Ly$\alpha$, 
the \ion{C}{IV}\,$\lambda$1549 profile has a shoulder-like feature and the 
\ion{He}{II}\,$\lambda$1640 profile has a narrow emission feature. The coincidence between 
the peak of Ly$\alpha$ and that of the non-resonant \ion{He}{II}\,$\lambda$1640 suggests that the 
absorption on Ly$\alpha$ is not strong, and the emission at zero velocity is not absorbed. This 
supports our assertion that the peak of Ly$\alpha$ is a good tracer of the systemic velocity. 
As narrow lines' contribution to the total \ion{C}{IV}\,$\lambda$1549 profile is weaker and is a 
blended doublet, we cannot assess whether the absorption feature in the narrow Ly$\alpha$ 
is present in \ion{C}{IV}\,$\lambda$1549. To measure and remove the narrow line contributions in 
\ion{C}{IV}\,$\lambda$1549 and \ion{He}{II}\,$\lambda$1640 we need to estimate the broad-line 
profiles near them, although our goal is not to extract the broad lines. We speculate that the 
narrow emission lines all arise on spatial scales much further out than the quasar broad-line 
region. We discuss evidence supporting this in Section 3.4.

We fit a single power law to line-free regions. 
This continuum fit is subtracted from the data when analyzing the 
line emissions. 

The \ion{C}{IV}\,$\lambda$1549 emission profile is broad and asymmetric. This emission is a 
doublet. In the panel for \ion{C}{IV}\,$\lambda$1549 in Fig.~\ref{fig:fitcorespec}, velocities 
are defined relative to the average of the rest wavelengths of the doublet, 1549.48\,\AA. 
The doublet separation of \ion{C}{IV}\,$\lambda\lambda$1548,1550 at 498\,km\,s$^{-1}$ 
is much smaller than the velocity width of the broad component of one line of this doublet, but is 
comparable to the velocity width of the narrow component of one line of this doublet. The 
shoulder-like feature is on the red side of the emission profile. A single Gaussian function can 
capture the shape of the red side of the broad component of the total blended doublet, while two 
Gaussians are necessary to capture the shape of the narrow component of the total blended doublet. 
We thus fit the \ion{C}{IV}\,$\lambda$1549 region with a broad Gaussian function plus two narrow 
Gaussian functions. The two narrow Gaussian functions are fixed at the doublet separation, and 
represent a \ion{C}{IV}\,$\lambda\lambda$1548,1550 doublet whose velocity centroids and widths are 
fixed at the narrow Ly$\alpha$'s values. 
With respect to the average rest wavelength of the doublet, i.e.\ zero velocity is midway between 
the doublet lines, on Fig.~\ref{fig:fitcorespec} the two narrow Gaussians at zero velocity with 
respect to the systemic are placed at $-249\,{\rm km\,s^{-1}}$ and $+249\,{\rm km\,s^{-1}}$. 
During the fitting process we mask the asymmetric blue side. 


The \ion{He}{II}\,$\lambda$1640 emission is broad and asymmetric. The narrow component is 
on the red side of this line profile. We fit with a broad Gaussian function plus a narrow Gaussian 
function. We fix the narrow Gaussian's velocity centroid and width at the narrow Ly$\alpha$'s 
values. 
During the fitting we mask the asymmetric blue side. 

As illustrated in Fig.~\ref{fig:fitcorespec}, the narrow emission in all of Ly$\alpha$, 
\ion{C}{IV}\,$\lambda\lambda$1548,1550, and \ion{He}{II}\,$\lambda$1640 is well fitted with a 
single component. All three narrow emission lines have similar kinematics. The results of the 
above line fitting processes 
are reported in Table~\ref{tab:fitcoreintspecs}. We present formal $\chi^2$ errors on parameters 
that are free during the fitting, and we note that the formal $\chi^2$ values may underestimate 
the true uncertainties due to degeneracies in the parameters. In the same table we also present 
fitting results on the spatially-integrated, halo-scale emissions that are described in Section 
3.4.1. 

\begin{table}
\caption{Results of Gaussian Fitting the Narrow Emissions in the Central 1-arcsec and the Spatially-Integrated Halo Emissions}
\label{tab:fitcoreintspecs}
\setlength{\tabcolsep}{0.07in}
\begin{tabular}{lccc}
\hline
Line Component & Flux & Centre\tablenotemark{a} & 1$\sigma$ Dispersion \\
 & (erg\,s$^{-1}$\,cm$^{-2}$) & (km\,s$^{-1}$) & (km\,s$^{-1}$) \\
\hline
\multicolumn{1}{l}{} & \multicolumn{3}{c}{Narrow Emissions in Central Spectrum} \\
\hline
\ion{H}{I} Ly$\alpha$ & $(1.72\pm0.05)\times10^{-16}$ & $0\pm8$ & $253\pm5$ \\
\ion{C}{IV}\,$\lambda$1548 & $(6.4\pm0.9)\times10^{-18}$ & $0$ & $253$ \\
\ion{C}{IV}\,$\lambda$1550\tablenotemark{b} & $6.4\times10^{-18}$ & $0$ & $253$ \\
\ion{He}{II}\,$\lambda$1640 & $(1.1\pm0.1)\times10^{-17}$ & $0$ & $253$ \\
\hline
\multicolumn{1}{l}{} & \multicolumn{3}{c}{Spatially-Integrated Halo Emissions} \\
\hline
\ion{H}{I}\,Ly$\alpha$ & $(1.45\pm0.01)\times10^{-15}$ & $-46\pm3$ & $304\pm2$ \\
\ion{C}{IV}\,$\lambda$1548 & $(3.41\pm0.07)\times10^{-17}$ & $-44\pm3$ & $319\pm4$ \\
\ion{C}{IV}\,$\lambda$1550\tablenotemark{b} & $3.41\times10^{-17}$ & $-44$ & $319$\\
\ion{He}{II}\,$\lambda$1640 & $(1.78\pm0.07)\times10^{-17}$ & $-45\pm7$ & $238\pm7$ \\
\hline
\end{tabular}
\tablenotetext{a}{\raggedright The narrow Ly$\alpha$ line centre in the central spectrum is defined to be the zero velocity. Other lines in the central spectrum are forced to be at zero velocity.}
\tablenotetext{b}{\raggedright \ion{C}{IV}\,$\lambda$1550 is locked to \ion{C}{IV}\,$\lambda$1548.}
\end{table}

\section{Analysis and Results}

\subsection{Quasar Spectral Properties} 

A BOSS spectrum of the quasar was obtained on 2011 December 17, with a fibre aperture of 2-arcsec 
diameter. A 2-arcsec aperture spectrum extracted from the KCWI data has similar spectral 
resolution but much higher S/N than BOSS. The higher S/N reveals for the first time the presence 
of narrow emission components in \ion{C}{IV}\,$\lambda\lambda$1548,1550 and 
\ion{He}{II}\,$\lambda$1640, which have similar kinematics as the Ly$\alpha$ spike that was 
already clearly present in the BOSS spectrum, as shown in Fig.~\ref{fig:KCWIBOSS}. 
The broad emission lines are spatially unresolved, hence we may compare emission line properties 
of a central 1-arcsec spectrum extracted from KCWI data to the BOSS spectrum. These observations 
taken two years apart in quasar's frame also provide a test for variability. 

The $z_{\rm sys,Ly\alpha}=2.3184$ we find from the Ly$\alpha$ spike is very similar to the 
$z_{\rm sys,[OIII]}=2.3198$ measured by \cite{Perrotta+19} from the narrow component in the 
[\ion{O}{III}]\,$\lambda\lambda$4959,5007 emission lines. The two systemic redshift values are 
consistent with each other and within one spectral resolution element $\approx$150\,km\,s$^{-1}$.
The velocity width of the narrow Ly$\alpha$, $\sigma_{\rm Ly\alpha}=253$\,km\,s$^{-1}$ is also 
similar to the velocity width of the narrow [\ion{O}{III}]\,$\lambda\lambda$4959,5007, 
$\sigma_{\rm [OIII]}=307$\,km\,s$^{-1}$, implying a possible origin in the same spatial region. 

Our measurements made on the central 1-arcsec KCWI spectrum are reported in 
Table~\ref{tab:directqsospec}. 
We measure the total blended doublet \ion{Si}{IV}\,$\lambda$1398, the total and the broad 
component of the blended doublet \ion{C}{IV}\,$\lambda$1549, and the total and the broad component 
of the singlet \ion{He}{II}\,$\lambda$1640. We isolate the broad components by subtracting 
the narrow emission line fits from the data. 
For the total \ion{Si}{IV}\,$\lambda$1398 emission, the total \ion{C}{IV}\,$\lambda$1549 emission, 
and the broad component of \ion{C}{IV}\,$\lambda$1549, we report the velocity centroid, FWHM 
velocity, the kurtosis index $kt_{80}$, and the rest equivalent width values. 
As the total \ion{C}{IV}\,$\lambda$1549 line profile is dominated by the broad component, the 
measurements on the broad component are very similar to the measurements on the total line profile. 
The total \ion{He}{II}\,$\lambda$1640 emission is weak and has contribution from narrow emission 
that is comparable to the broad emission. We do not attempt to quantify the kurtosis of the total 
\ion{He}{II}\,$\lambda$1640 or the broad component of \ion{He}{II}\,$\lambda$1640. 
The $kt_{80}$ quantity characterizes line profile shapes and measures the velocity width at 
80 per cent of the peak height divided by the width at 20 per cent. For reference, a single 
Gaussian function has $kt_{80}=0.372$ while most quasars have substantial line wings yielding 
lower values 0.15\textendash0.3. 

Our measurements on the total \ion{C}{IV}\,$\lambda$1549 profile in the central 1-arcsec KCWI 
spectrum are similar to the measurements by \cite{Hamann+17} on the BOSS spectrum. They report a 
blueshift of 2267\,km\,s$^{-1}$ estimated from the profile centre at half maximum relative to the 
Ly$\alpha$ spike, a rest equivalent width of 107$\pm$6\,\AA, a FWHM velocity of 
4540$\pm$200\,km\,s$^{-1}$, and a $kt_{80}$ of 0.37. Small differences between these measurements 
and Table~\ref{tab:directqsospec} may be due to quasar variability or differences in measurement 
techniques. The broad Ly$\alpha$-\ion{N}{V}\,$\lambda$1240 complex has decreased in strength since 
BOSS, although we do not attempt to measure the rest equivalent widths of these heavily blended 
lines. 
The overall similarity between the two sets of measurements indicates that the peculiar spectral 
properties of this ERQ persisted over this period of two years in the quasar rest frame, 
consistent with the repeat observations of other ERQs in \cite{Hamann+17}.
\begin{table}
\caption{{\bf Quasar Emission Line Properties in the Central 1\,arcsec}}
\label{tab:directqsospec}
\setlength{\tabcolsep}{0.06in}
\begin{tabular}{lcccc}
\hline
Line Component & Centroid\tablenotemark{a} & FWHM & $kt_{80}$\tablenotemark{b} & REW\tablenotemark{c} \\
 & (km\,s$^{-1}$) & (km\,s$^{-1}$) & & (\AA) \\
\hline
Total \ion{Si}{IV}\,$\lambda$1398 & $-1770\pm180$ & $4750\pm280$ & $0.37\pm0.02$ & $30\pm1$ \\
Total \ion{C}{IV}\,$\lambda$1549 & $-2180\pm50$ & $5110\pm60$ & $0.30\pm0.01$ & $102\pm1$ \\
Broad \ion{C}{IV}\,$\lambda$1549\tablenotemark{d} & $-2240\pm50$ & $4880\pm70$ & $0.31\pm0.01$ & $99\pm1$ \\
Total \ion{He}{II}\,$\lambda$1640 & $-2110\pm250$ & $3431\pm120$ & \textendash & $19\pm1$ \\
Broad \ion{He}{II}\,$\lambda$1640\tablenotemark{d} & $-2430\pm300$ & $2330\pm140$& \textendash & $16\pm1$ \\
\hline
\end{tabular}
\tablenotetext{a}{\raggedright Centroid of the line or doublet in velocity relative to the systemic redshift.}
\tablenotetext{b}{\raggedright Kurtosis index $kt_{80}$ of the line or doublet.}
\tablenotetext{c}{\raggedright Equivalent width of the line or doublet in rest-frame wavelength.}
\tablenotetext{d}{\raggedright The broad component is obtained from subtracting the narrow emission line fits from the data.}
\end{table}

\subsection{Morphology and Brightness of the Extended Line Emissions}

\subsubsection{Optimal Extraction of Diffuse Emissions}

Relative to standard pseudo-narrowband images, an optimal extraction algorithm better captures the 
morphology and kinematic features of diffuse extended line emissions. We optimally extract 
extended line emissions, if any, in the \ion{H}{I}\,Ly$\alpha$, \ion{C}{IV}\,$\lambda$1549, and 
\ion{He}{II}\,$\lambda$1640, follow an algorithm similar to that described in \cite{Borisova+16a}, 
\cite{ArrigoniBattaia+19}, \cite{Cai+19}, and \cite{Farina+19}. We subtract the quasar point 
spread function constructed using the unresolved quasar spectral template. We only use portions of 
the cube within $\pm5000$\,km\,s$^{-1}$ around the line wavelengths at the systemic redshift. This 
relative velocity range corresponds to $\sim$150\,\AA\ in observer's frame. Each wavelength layer 
of the data cube and the variance cube is smoothed with a two-dimensional Gaussian kernel of 
$\sigma_{\rm spatial}=0.4$\,arcsec. The FWHM of the chosen kernel, 0.9\,arcsec, matches the seeing 
disc and is conservative compared to the size of the point spread function measured on the data. 
The smoothing in spatial dimensions without smoothing in wavelength helps reveal extended and 
narrow features. We then create a three-dimensional segmentation mask that selects connected 
voxels (volume pixels) around a given line wavelength that are of S/N $>2$ and in groups of 
minimal 200 voxels. 
In three dimensions, connected voxels are neighbours to every voxel that touches one of their 
faces, edges, or corners. The minimum number of connected voxels is empirically derived to avoid 
selecting instrument artifacts that are not fully removed in data reduction and post-processing. 
For each of the Ly$\alpha$, \ion{C}{IV}\,$\lambda$1549, and \ion{He}{II}\,$\lambda$1640, an 
optimally extracted image is constructed by summing the flux along the wavelength direction for 
only the voxels selected by the line's three-dimensional segmentation mask. 

\subsubsection{Morphology and Brightness Measurements}

\begin{figure*}
\includegraphics[width=0.36\textwidth]{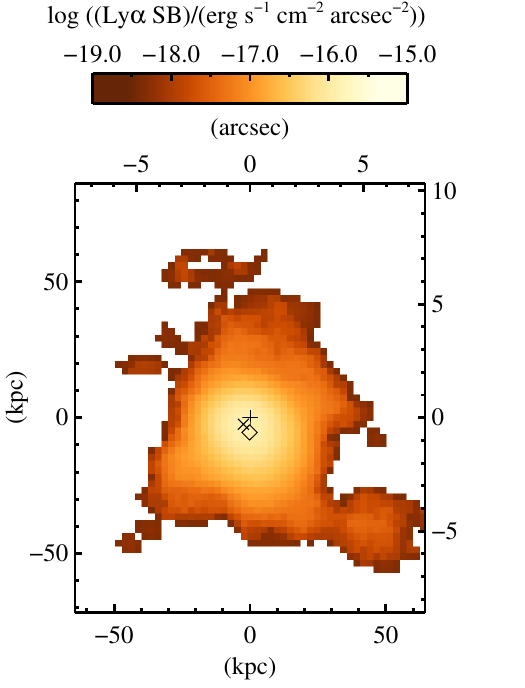}
\hspace{-0.3in}
\includegraphics[width=0.36\textwidth]{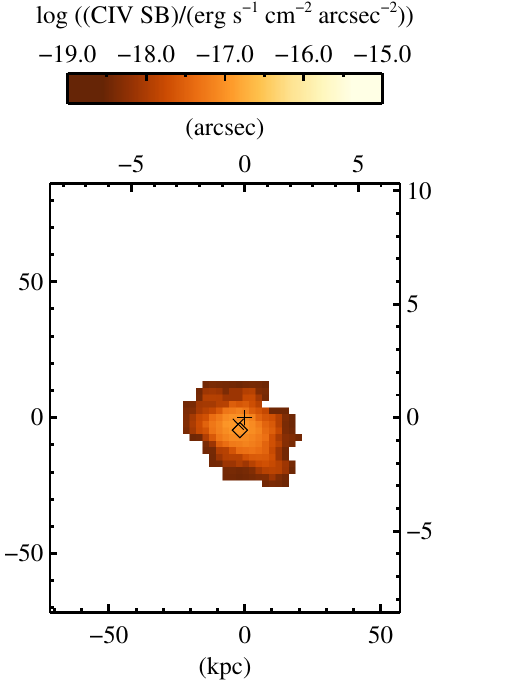}
\hspace{-0.4in}
\includegraphics[width=0.36\textwidth]{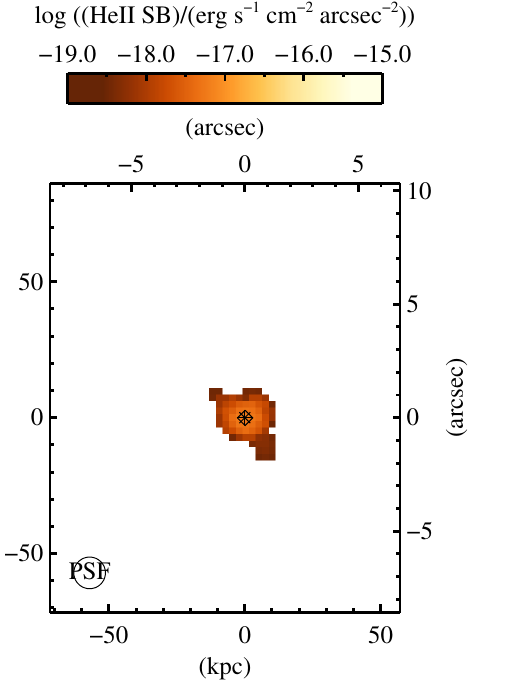}
\caption{Optimally extracted images of Ly$\alpha$, \ion{C}{IV}\,$\lambda$1549, and 
\ion{He}{II}\,$\lambda$1640 after subtracting the quasar. On each image the white contours 
indicate the area detected above a S/N threshold of two. On each image the plus symbol marks the 
position of the quasar, the cross symbol marks the position of the peak of the extended line 
emission, and the diamond symbol marks the position of the centroid of the extended line emission. 
The spatial axes of each image have angular distances transformed to projected proper distances. 
On the \ion{He}{II}\,$\lambda$1640 image, the FWHM of the empirical point spread function is 
displayed. 
On these maps and all subsequent figures of maps, north is pointing upright.}
\label{fig:optextracts}
\end{figure*}

In Fig.~\ref{fig:optextracts} we show the optimally extracted images of the Ly$\alpha$,  
\ion{C}{IV}\,$\lambda$1549, and \ion{He}{II}\,$\lambda$1640 line emissions down to a S/N 
threshold of two, corresponding to a surface brightness limit 
$\approx\!3\times10^{-19}$\,erg\,s$^{-1}$\,cm$^{-2}$\,arcsec$^{-2}$. The spatially-integrated 
luminosities, maximum linear sizes, and areas covered by the line emissions are tabulated in 
Table~\ref{tab:morph}. The Ly$\alpha$ and \ion{C}{IV}\,$\lambda$1549 line emissions are clearly 
spatially extended, and their maximum linear sizes within covered areas of S/N $>2$ are 
respectively 140\,kpc and 47\,kpc. The \ion{He}{II}\,$\lambda$1640 emission is marginally 
spatially resolved with a maximum linear size of 33\,kpc, and the majority of the line flux is 
contained within the FWHM of the point spread function that is 1.4\,arcsec or 12\,kpc.  

We define an inner halo region $\sim$(20\textendash30)\,kpc from the centre and an outer halo 
region beyond. This definition is motivated by the morphology analysis, the line ratio analysis, 
and the kinematics analysis in this and the next two subsections, where we find that measured 
properties transition near these distances. Evidence that gaseous halo properties transition near 
similar scales has been reported in the literature and is discussed in Section~4.2. 

We quantify the spatial asymmetry of the flux distributions in the optimally extracted images as 
follows. For each line, we calculate the halo centroid as the first spatial moment of the line 
flux within areas of S/N $>2$. We then measure the projected distance between the quasar and the 
peak of the line emission, and the projected distance between the quasar and the centroid. We 
then calculate two dimensionless parameters measuring circular asymmetry that respectively 
emphasize the brighter inner halo region and the overall diffuse halo emission. These quantities 
are tabulated in Table~\ref{tab:morph}.  

The uncertainty in the projected distance between the quasar and an extended line emission's peak 
or centroid is estimated to be about half a spaxel's size, or $\approx$1\,kpc in physical distance. 
The peaks of Ly$\alpha$ and \ion{C}{IV}\,$\lambda$1549 are spatially offset from the position of 
the quasar by similar amounts of 3\,kpc and in similar general directions. The centroids of 
Ly$\alpha$ and \ion{C}{IV}\,$\lambda$1549 are spatially offset from the position of the quasar by 
similar amounts of 5\,kpc and in similar general directions. 
The extended emission in all three lines have their peaks and centroids in close proximity of the 
quasar, indicating only a low level of asymmetry in the inner halo. The position of the Ly$\alpha$ 
halo centroid is used in assessing spatial asymmetry in this section and the position of the 
Ly$\alpha$ halo peak is used in analysis of aperture kinematics in Section~3.4.2. 

We define an elliptical eccentricity parameter $e_{\rm weight}$ using flux-weighted second-order 
spatial moments with respect to the halo centroid according to the formulae in 
\cite{OSullivanChen20}. This parameter reflects the ratio of the semiminor axis to the semimajor 
axis of the extended line emission. A value of $e\ll1$ indicates very circular morphology, while 
$e\approx1$ indicates very asymmetric shape. The definition of $e_{\rm weight}$ tends to weight 
toward asymmetries on smaller scales and with respect to where most of the halo emission is. This 
$e_{\rm weight}$ parameter is related to the flux-weighted asymmetry parameter $\alpha$ defined in 
\cite{ArrigoniBattaia+19} or \cite{Cai+19} by $e_{\rm weight}=\sqrt{1-\alpha_{\rm weight}^2}$. We 
define another elliptical eccentricity parameter $e_{\rm unweight}$ using flux-unweighted 
second-order spatial moment with respect to the quasar position. The definition of 
$e_{\rm unweight}$ better characterizes asymmetries on larger scales and with respect to the 
powering source, the quasar. This $e_{\rm unweight}$ parameter is related to the flux-unweighted 
asymmetry parameter $\alpha$ defined in \cite{denBrok+20} or \cite{Mackenzie+21} by 
$e_{\rm unweight}=\sqrt{1-\alpha_{\rm unweight}^2}$. For the optimally extracted Ly$\alpha$ halo, 
we find $e_{\rm weight}=0.44$ and $e_{\rm unweight}=0.69$. The relatively low value of 
$e_{\rm weight}$ indicates that the halo is circularly symmetric in regions where most of the line 
flux is contained. Indeed, the halo appears symmetric in the inner region on scales 
$\lesssim$30\,kpc from the centroid. With a relatively low value of $e_{\rm weight}$, the higher 
value of $e_{\rm unweight}$ indicates some level of asymmetry on larger scales. This large-scale 
asymmetry characterizes the filamentary appearance of the halo in the outer region on 
$\gtrsim$30\,kpc scales. 

The Ly$\alpha$ and \ion{C}{IV}\,$\lambda$1549 emissions are spatially extended enough 
to warrant examination of their radial profiles. While optimally extracted line images better 
capture the morphology and kinematic features, for the purpose of calculating surface brightness 
radial profiles we use fixed width pseudo-narrowband images sliced from the quasar-subtracted, 
unsmoothed data cube. The choice of a fixed width pseudo-narrowband recovers all possible fluxes in 
extended regions, allows uniform comparison of detections and non-detections, and matches the 
methods in literature works for comparison. The widths of the Ly$\alpha$ narrowband 
and the \ion{C}{IV}\,$\lambda$1549 narrowband are set at $\pm1000$\,km\,s$^{-1}$ from the line 
rest wavelengths. We sum these two slices of the cube in the spectral dimension. We then bin 
the pseudo-narrowband images in concentric annuli that are centred on the quasar's position and 
have bin widths uniformly spaced in a logarithmic scale. Then in each annular bin we circularly 
average the surface brightness. For comparison with profiles of other line-emitting haloes in the 
literature at different redshifts, we multiply all surface brightness values by their cosmological 
dimming factor $(1+z)^4$. For each annular bin, we calculate surface brightness limit under the 
assumption that there are no unaccounted residuals from the sky, scattered light, continuum 
sources, or imperfect calibrations, or unaccounted covariances among spaxels. 
We sum the respective slice of the variance cube in the spectral dimension. We then sum the 
spaxels belonging to that annular aperture, take the square root, and divide by the angular area 
of that annulus. The expected surface brightness limit of an annular bin thus scales with the 
inverse square root of the annular aperture area. 

\begin{figure}
\includegraphics[width=\columnwidth]{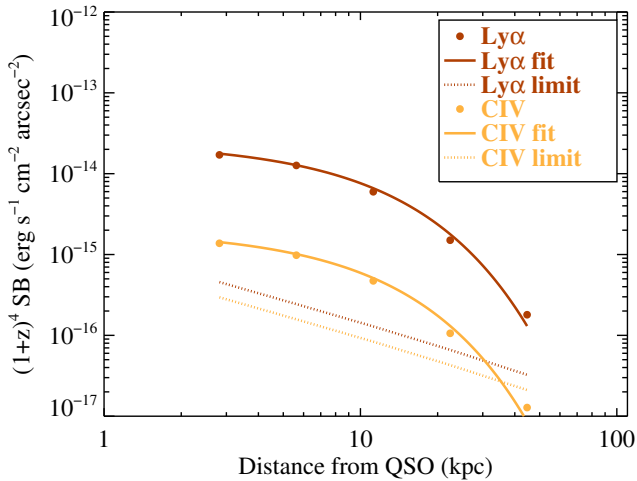}
\caption{Circularly averaged surface brightness radial profiles of Ly$\alpha$ and
\ion{C}{IV}\,$\lambda$1549, created from their pseudo-narrowband images. Bins of surface
brightness are measured in concentric annuli centred on the quasar, and the bin widths are
uniformly spaced on a logarithmic scale. Brown colours show Ly$\alpha$ and orange colours show
\ion{C}{IV}\,$\lambda$1549. The points are the measured data. The solid curves are exponential 
function fits to the data points. The dotted lines are the pseudo-narrowband 2$\sigma$ surface 
brightness limits in the annular apertures. All surface brightnesses are corrected for 
cosmological dimming.}
\label{fig:SBradial}
\end{figure}

Fig.~\ref{fig:SBradial} shows the surface brightness radial profiles and the 2$\sigma$ 
surface brightness limit profiles of Ly$\alpha$ and \ion{C}{IV}\,$\lambda$1549. 
Table~\ref{tab:SBradial} presents the data of these profiles. We find 
that the surface brightness radial profiles are better described by exponential laws than power 
laws. To each of Ly$\alpha$ and \ion{C}{IV}\,$\lambda$1549 we fit the function 
${\rm SB}_{{\rm Ly}\alpha}(r)=C_e\exp{(-r/r_h)}$, where $C_e$ is the normalization, $r$ is the 
projected distance from the quasar, and $r_h$ is the scale length. The fits are shown 
in Fig.~\ref{fig:SBradial}. The resulting parameters for Ly$\alpha$ are 
$C_e=(2.46\pm0.04)\times10^{-14}$\,erg\,s$^{-1}$\,cm$^{-2}$\,arcsec$^{-2}$ and 
$r_h=(8.5\pm0.1)$\,kpc, which describe a centrally concentrated profile. The resulting parameters 
for \ion{C}{IV}\,$\lambda$1549 are 
$C_e=(2.0\pm0.2)\times10^{-15}$\,erg\,s$^{-1}$\,cm$^{-2}$\,arcsec$^{-2}$ and 
$r_h=(8.2\pm0.7)$\,kpc. 

\begin{table*}
\caption{Size and Morphology Measures of the Extended Line Emissions}
\label{tab:morph}
\setlength{\tabcolsep}{0.1in}
\begin{tabular}{lcccccccc}
\hline
Line Name & Area Covered & Maximum Linear Size & Luminosity & Peak to QSO & Centroid to QSO & $e_{\rm weight}$\tablenotemark{c} & $e_{\rm unweight}$\tablenotemark{d} \\
 & & & & Distance\tablenotemark{a} & Distance\tablenotemark{b} & & \\
 & (kpc$^2$) & (kpc) & (erg\,s$^{-1}$) & (kpc) & (kpc) & & \\
\hline
\ion{H}{I}\,Ly$\alpha$ & 7005 & 140 & 5.11$\times10^{43}$ & 3$\pm$1 & 5$\pm$1 & 0.44 & 0.69 \\
\ion{C}{IV}\,$\lambda$1549 & 1210 & 47 & 1.81$\times10^{42}$ & 3$\pm$1 & 5$\pm$1 & 0.71 & 0.64 \\
\ion{He}{II}\,$\lambda$1640 & 437 & 33 & 4.90$\times10^{41}$ & 0$\pm$1 & 0$\pm$1 & \textendash & \textendash \\
\hline
\end{tabular}
\tablenotetext{a}{\raggedright Distance between the position of the peak of halo emission and the position of the quasar.}
\tablenotetext{b}{\raggedright Distance between the position of the centroid of halo emission and the position of the quasar.}
\tablenotetext{c}{\raggedright Flux-weighted elliptical eccentricity with respect to the halo centroid position $e_{\rm weight}$, calculated from flux-weighted spatial moments.}
\tablenotetext{d}{\raggedright Unweighted elliptical eccentricity with respect to the quasar position $e_{\rm unweight}$, calculated from unweighted spatial moments.}
\end{table*}

\begin{table*}
\caption{Circularly Averaged Surface Brightness Profiles of the Extended Line Emissions}
\label{tab:SBradial}
\setlength{\tabcolsep}{0.1in}
\begin{tabular}{lcccc}
\hline
Radii\tablenotemark{a} & \ion{H}{I} Ly$\alpha$\tablenotemark{b} & Ly$\alpha$ 2$\sigma$ Limit\tablenotemark{c} & \ion{C}{IV}\,$\lambda$1549\tablenotemark{b} & \ion{C}{IV} 2$\sigma$ Limit\tablenotemark{c} \\
\multicolumn{1}{l}{(kpc)} & \multicolumn{4}{c}{(erg\,s$^{-1}$\,cm$^{-2}$\,arcsec$^{-2}$)} \\
\hline
2\textendash4 & $(1708.6\pm35.1)\times10^{-17}$ & $45.6\times10^{-17}$&$(137.7\pm18.4)\times10^{-17}$ & $29.6\times10^{-17}$ \\
4\textendash8 & $(1264.9\pm16.6)\times10^{-17}$ & $24.4\times10^{-17}$&$(98.1\pm9.1)\times10^{-17}$ & $15.8\times10^{-17}$ \\
8\textendash16 & $(597.0\pm7.4)\times10^{-17}$ & $12.9\times10^{-17}$&$(47.2\pm4.4)\times10^{-17}$ & $8.4\times10^{-17}$ \\
16\textendash32 & $(150.3\pm3.5)\times10^{-17}$ & $6.6\times10^{-17}$&$(10.6\pm2.2)\times10^{-17}$ & $4.3\times10^{-17}$ \\
32\textendash63 & $(18.0\pm1.7)\times10^{-17}$ & $3.2\times10^{-17}$&$(1.3\pm1.1)\times10^{-17}$ & $2.1\times10^{-17}$ \\
\hline
\end{tabular}
\tablenotetext{a}{\raggedright Radii spanned by the annular bins.}
\tablenotetext{b}{\raggedright Corrected for cosmological dimming.}
\tablenotetext{c}{\raggedright Expected 2$\sigma$ surface brightness limit in the aperture, under the assumptions of perfect calibrations and other processings of the data.}
\end{table*}

\subsection{Line Ratios in the Extended Emissions} 

\begin{figure*}
\includegraphics[width=0.36\textwidth]{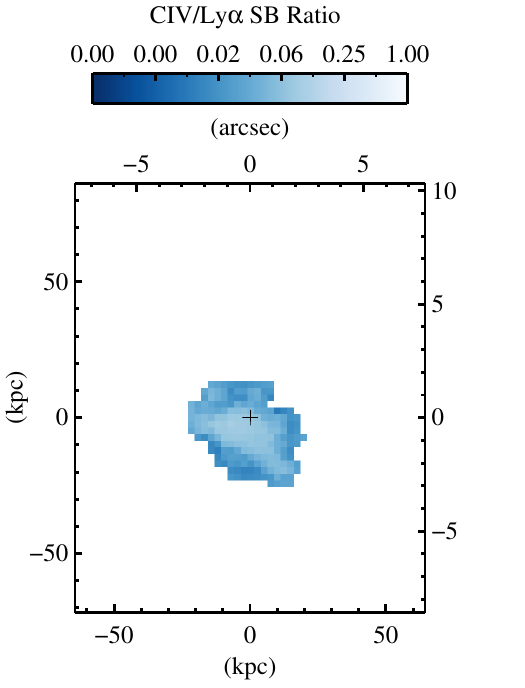}
\hspace{-0.3in}
\includegraphics[width=0.36\textwidth]{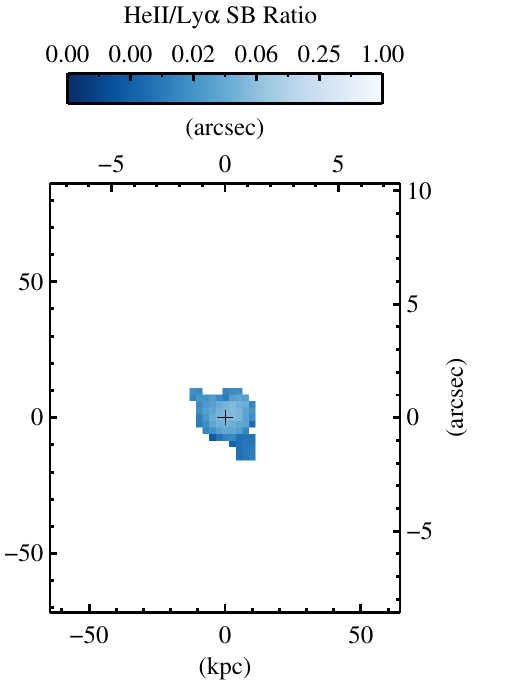}
\caption{Maps of the \ion{C}{IV}\,$\lambda$1549/Ly$\alpha$ line brightness ratio, and 
the \ion{He}{II}\,$\lambda$1640/Ly$\alpha$ line brightness ratio. The ratio values are
calculated only for regions where Ly$\alpha$ and \ion{C}{IV}\,$\lambda$1549 are both detected, and 
regions where Ly$\alpha$ and \ion{He}{II}\,$\lambda$1640 are both detected. The plus symbol marks 
the position of the quasar.}
\label{fig:ratiomaps}
\end{figure*}

We construct maps of surface brightness ratios from the optimally extracted line images of 
\ion{H}{I}\,Ly$\alpha$, \ion{C}{IV}\,$\lambda$1549, and \ion{He}{II}\,$\lambda$1640. We compute 
the ratios only in regions where both lines in a ratio are detected at S/N $>2$. These regions 
are out to $\sim$20\,kpc from the quasar for \ion{C}{IV}\,$\lambda$1549/Ly$\alpha$ and out to 
$\sim$10\,kpc from the quasar for \ion{He}{II}\,$\lambda$1640/Ly$\alpha$. 
Fig.~\ref{fig:ratiomaps} shows maps of \ion{C}{IV}\,$\lambda$1549/Ly$\alpha$ and 
\ion{He}{II}\,$\lambda$1640/Ly$\alpha$. Table~\ref{tab:ratiomaps} presents the ratios 
of line fluxes spatially integrated over regions of significant detection. The same ratios may be 
measured from the extracted narrow line emissions from the 1-arcsec aperture spectrum centred on 
the quasar, and they are also presented in Table~\ref{tab:ratiomaps}. 

The central 1-arcsec \ion{C}{IV}\,$\lambda$1549/\ion{He}{II}\,$\lambda$1640 ratio is of order 
unity, while the spatially-integrated ratio shows the \ion{C}{IV}\,$\lambda$1549 brightness is 
somewhat higher than \ion{He}{II}\,$\lambda$1640. Quasar photoionization models predict 
\ion{C}{IV}\,$\lambda$1549 and \ion{He}{II}\,$\lambda$1640 brightnesses of the same order for 
the ranges of metallicities and gas densities in circumgalactic environments 
\citep{FeltreCharlotGutkin16,Humphrey+19}. The observed line ratios are consistent with quasar 
photoionization being the major powering mechanism of the extended line emissions. 

Both \ion{C}{IV}\,$\lambda$1549/Ly$\alpha$ and \ion{He}{II}\,$\lambda$1640/Ly$\alpha$ ratios 
decline outwardly. This reflects a steeper radial decline in \ion{C}{IV}\,$\lambda$1549 brightness 
and \ion{He}{II}\,$\lambda$1640 brightness than Ly$\alpha$. The spatially-integrated values of 
\ion{C}{IV}\,$\lambda$1549/Ly$\alpha$ and \ion{He}{II}\,$\lambda$1640/Ly$\alpha$ are lower than 
the central 1-arcsec line ratios, again reflecting the outward declines. The outward decline of 
these line ratios reflects a decline of the ionization parameter with distance from the ionizing 
source \citep{FeltreCharlotGutkin16,Humphrey+19}, which holds for a density gradient less steep   
than inverse distance squared. 

The \ion{He}{II}\,$\lambda$1640/Ly$\alpha$ ratios at all significantly detected spaxels are much 
lower than the theoretical limits for Case A or Case B recombination in a fully photoionized gas, 
which are respectively 0.23 and 0.3 \citep{Cantalupo+19}. This suggests that helium is not fully 
doubly ionized, and the amount of \ion{He}{II}\,$\lambda$1640 emission by fluorescent 
recombination is ionization-bounded. 


Knowledge of both \ion{C}{IV}\,$\lambda$1549/\ion{He}{II}\,$\lambda$1640 and 
\ion{C}{III}]\,$\lambda$1909/\ion{C}{IV}\,$\lambda$1549 would have provided diagnostics for 
the ionization levels and the metallicities of the emitting gas, given quasar photoionization 
models. Our data do not cover \ion{C}{III}]\,$\lambda$1909, and the 
\ion{C}{IV}\,$\lambda$1549/\ion{He}{II}\,$\lambda$1640 ratio alone does not provide meaningful 
simultaneous constraints to the ionization parameter and the metallicity. In \cite{Guo+20} which 
study the average extended line emissions around $z\sim3$ luminous blue quasars, metallicities of 
circumgalactic gas at different ionization parameters are given on a line ratio diagnostic 
diagram. We may make the assumption that the ionization parameter $U$ of the inner halo region 
of J0006+1215 where both \ion{C}{IV}\,$\lambda$1549 and \ion{He}{II}\,$\lambda$1640 are detected 
is similar to that of \cite{Guo+20}. With this assumption, at $\log{U}\sim-1$ the central 1-arcsec 
\ion{C}{IV}\,$\lambda$1549/\ion{He}{II}\,$\lambda$1640 ratio of $\sim$1 corresponds to a 
metallicity $\sim$1\,${\rm Z_\odot}$, and the spatially-integrated line ratio of $\sim$3 
corresponds to a metallicity $\sim$0.5\,${\rm Z_\odot}$. This inner halo enrichment level is 
similar to the stacked inner halo of \cite{Guo+20}. The high enrichment level suggests 
centrally-driven outflow events of gas processed by multiple stellar generations and the outflows 
deposit materials out to at least $\sim$(20\textendash30)\,kpc from the host galaxy. 

\begin{table}
\caption{Summary of Line Ratios in the Extended Emission}
\label{tab:ratiomaps}
\setlength{\tabcolsep}{0.1in}
\begin{tabular}{lcc}
\hline
Lines & Inner Halo \tablenotemark{a} & Central 1\,arcsec \tablenotemark{b} \\
\hline
(\ion{C}{IV}\,$\lambda$1549)/(\ion{H}{I}\,Ly$\alpha$) & 0.04 & 0.07 \\
(\ion{He}{II}\,$\lambda$1640)/(\ion{H}{I}\,Ly$\alpha$) & 0.01 & 0.06 \\
(\ion{C}{IV}\,$\lambda$1549)(\ion{He}{II}\,$\lambda$1640) & 3.70 & 1.19 \\
\hline
\end{tabular}
\tablenotetext{a}{\raggedright Calculated in regions where both lines in a ratio have S/N $>$ 2.}
\tablenotetext{b}{\raggedright Calculated using extracted narrow emissions from the central 1\,arcsec.}
\end{table}

\subsection{Kinematics of the Extended Line Emissions}

We characterize the overall halo kinematics using spatially integrated and spatially resolved 
spectra. A baseline comparison for assessing the velocity fields would be the circular velocity of 
the host dark matter halo. J0006+1215 is selected from BOSS quasars. From clustering measurements 
of BOSS quasars at similar redshifts, the average host dark matter halo mass is 
$10^{12.4}h^{-1}$\,M$_\odot$ \citep{Eftekharzadeh+15}. At this dark matter halo mass and 
J0006+1215's redshift, and assuming a halo concentration parameter of 4, the maximum circular 
velocity is $v_{\rm circ}=338$\,km\,s$^{-1}$. From analytical arguments that are verified by 
numerical simulation results, the one-dimensional velocity dispersion is related to the circular 
velocity by $v_{\rm circ}\approx\sqrt{2}\sigma_{\rm 1D}$ 
\citep{BinneyTremaine08,LauNagaiKravtsov10,Munari+13}. The corresponding $\sigma_{\rm 1D}$ is thus 
239\,km\,s$^{-1}$. On the other hand, J0006+1215 has among the highest bolometric luminosities of 
BOSS quasars. Evidence has been found that obscured quasars and hyperluminous quasars on 
average reside in even more massive dark matter haloes typically 
$10^{13}h^{-1}$\,M$_\odot$ \citep{DiPompeo+17,Geach+19}. At this dark matter halo mass 
$v_{\rm circ}=536$\,km\,s$^{-1}$ and $\sigma_{\rm 1D}=379$\,km\,s$^{-1}$. Halo gas with measured 
line-of-sight speeds much greater than 379\,km\,s$^{-1}$ may thus be considered fast moving.  

\subsubsection{Spatially-Integrated Kinematics} 

\begin{figure}
\includegraphics[width=\columnwidth]{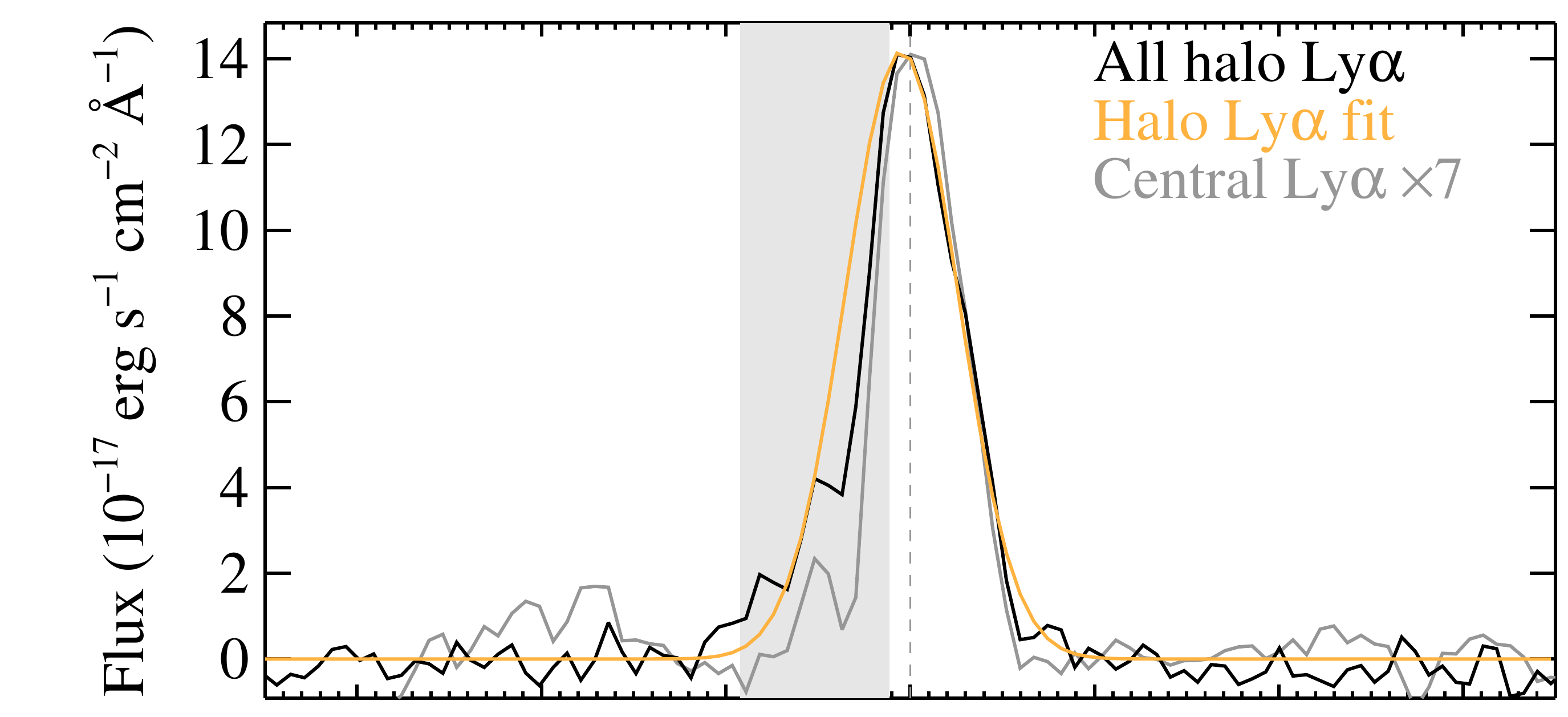}
\includegraphics[width=\columnwidth]{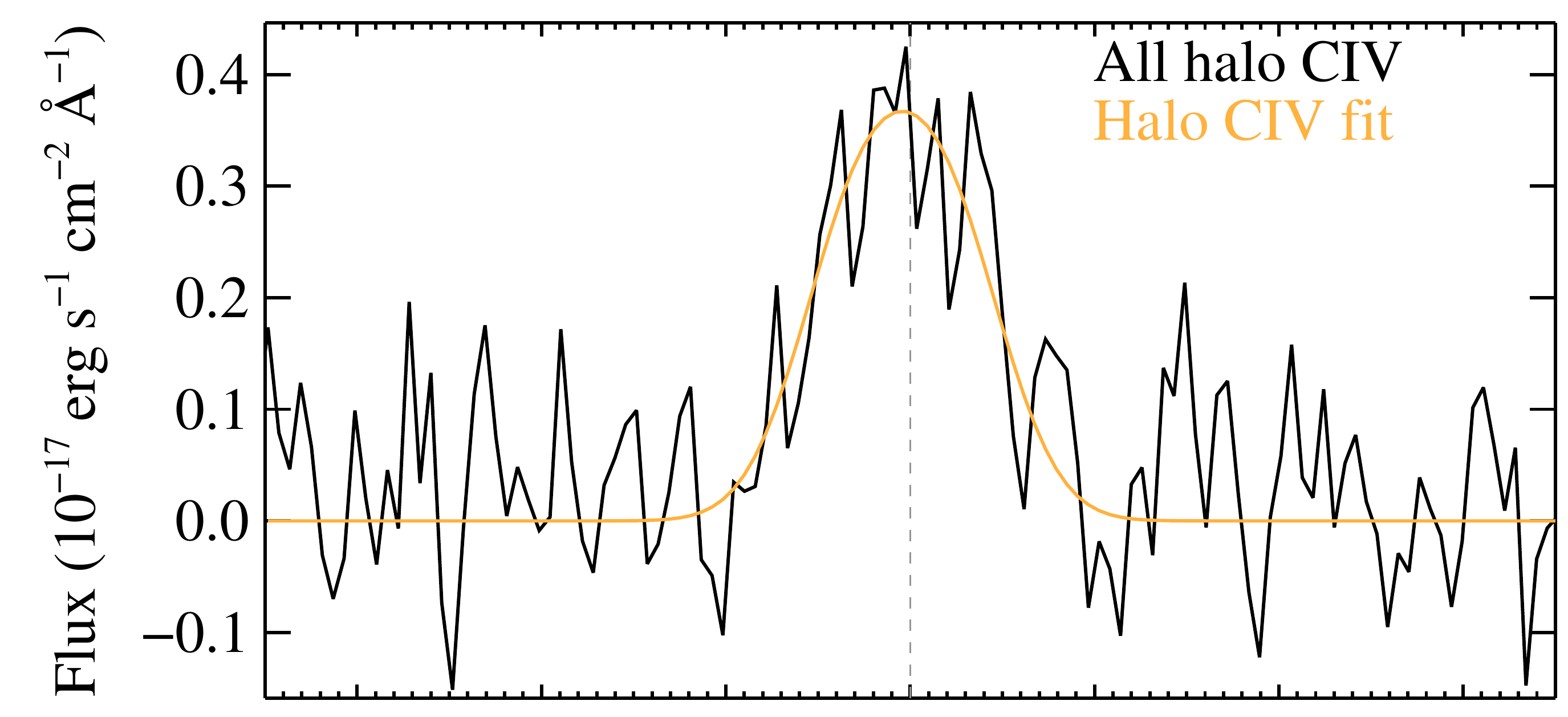}
\includegraphics[width=\columnwidth]{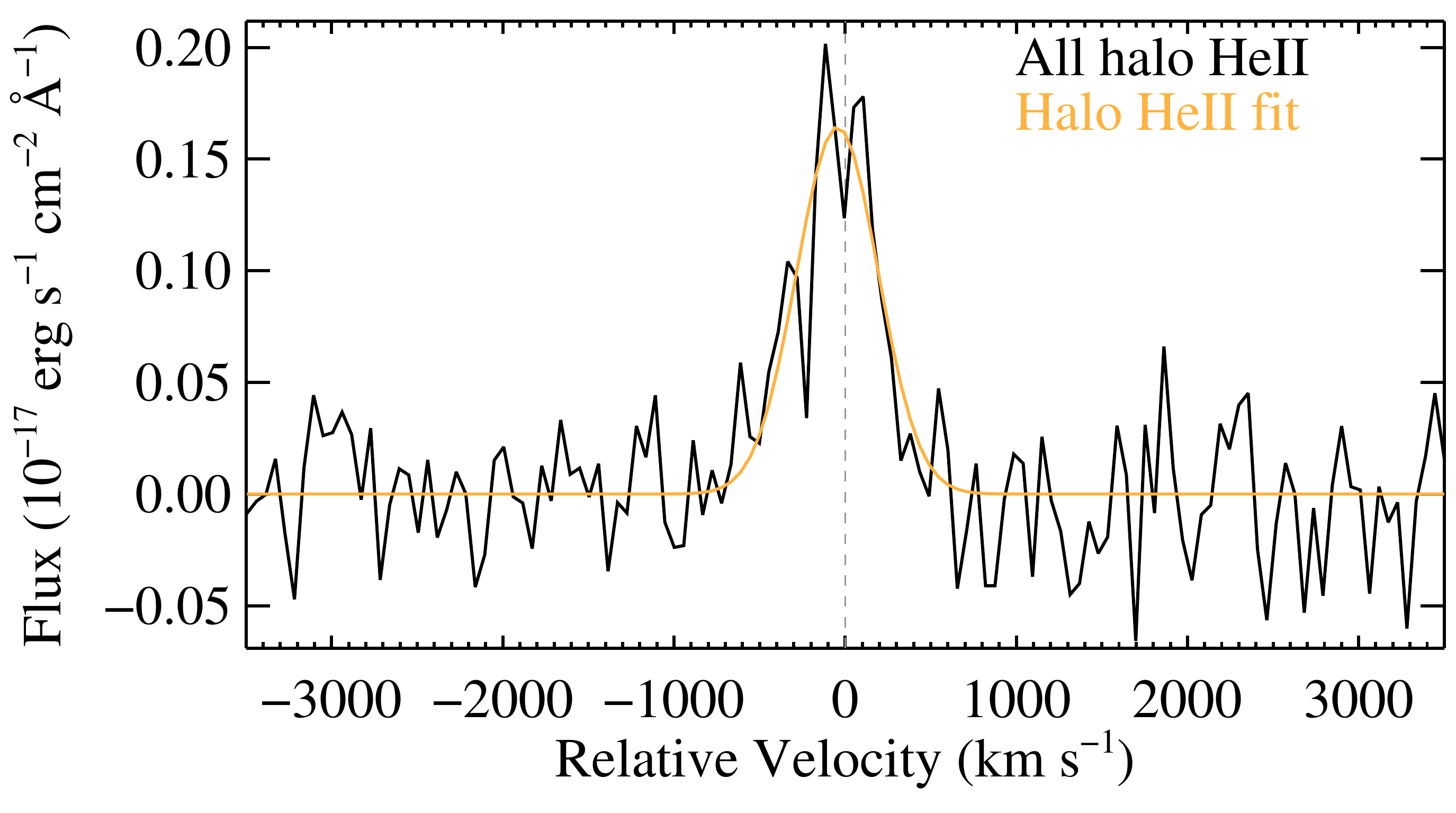}
\caption{Spatially-integrated, one-dimensional spectra of the extended line emissions. The spectra 
are extracted over areas covered by the three-dimensional segmentation masks collapsed in the 
spectral direction. Grey dashed lines mark locations of zero velocities. Top panel shows  
Ly$\alpha$. The black is the data. The orange curve is a single Gaussian fit to the data. During 
the fitting the absorption on the blue side of the line profile is masked, and is shaded in grey. 
The grey curve is the narrow Ly$\alpha$ emission region of a 1-arcsec aperture spectrum centred on 
the quasar minus a fit to the local pseudo-continuum underneath it. The peak flux of this central 
1-arcsec spectrum is rescaled to match the peak of the halo emission for comparison. 
Middle panel shows \ion{C}{IV}\,$\lambda\lambda$1548,1550. The orange curve is a tied double 
Gaussian fit to the data. Bottom panel shows \ion{He}{II}\,$\lambda$1640. The orange curve is a 
single Gaussian fit to the data.}
\label{fig:fitintspec}
\end{figure}

We extract the spatially-integrated spectra of Ly$\alpha$, \ion{C}{IV}\,$\lambda\lambda$1548,1550, 
and \ion{He}{II}\,$\lambda$1640. 
For each line or doublet, we select the spaxels for summing using a two-dimensional mask obtained 
by collapsing the spectral dimension of its three-dimensional segmentation mask. 
To describe the overall line profiles, we follow the procedure in Section 2.5 and 
Fig.~\ref{fig:fitcorespec} for fitting the narrow emission lines in the central 1-arcsec 
spectrum. 
The spatially-integrated emission-line spectra and our fits are presented in 
Fig.~\ref{fig:fitintspec}. The resulting velocity centres and Gaussian 1$\sigma$ velocity 
dispersions of the fits are presented in Table~\ref{tab:fitcoreintspecs}. 

In Section 2.5, we measure the redshift of the narrow Ly$\alpha$ emission in the central 1-arcsec 
spectrum and adopt it to be the systemic. The fitted line centres of spatially-integrated 
Ly$\alpha$, \ion{C}{IV}\,$\lambda\lambda$1548,1550, and \ion{He}{II}\,$\lambda$1640 are within one 
spectral resolution element from the systemic redshift. The similar line centres and line widths 
between the spatially-integrated emissions and the central 1-arcsec emissions support our 
assertion that the narrow emission lines in the central 1-arcsec spectrum originate far from the 
quasar at halo-scale distances. 

Ly$\alpha$ and \ion{C}{IV}\,$\lambda\lambda$1548,1550 are resonant lines. Resonant scattering is 
not efficient for \ion{C}{IV} photons, due to the much lower abundance of metals. 
\ion{He}{II}\,$\lambda$1640 and [\ion{O}{III}]\,$\lambda$5007 are non-resonant. The similar line 
profiles of Ly$\alpha$, \ion{He}{II}\,$\lambda$1640, and the narrow component of unresolved 
[\ion{O}{III}]\,$\lambda$5007 published in \cite{Perrotta+19} suggest that scattering of centrally 
produced Ly$\alpha$ photons is not a major contributor to the observed emission. 

The fitted 1$\sigma$ dispersion of spatially-integrated Ly$\alpha$ is somewhat broader than that 
of the central 1-arcsec narrow Ly$\alpha$. Likewise the fitted 1$\sigma$ dispersion of 
spatially-integrated \ion{C}{IV}\,$\lambda\lambda$1548,1550 is broader than the extracted narrow 
\ion{C}{IV}\,$\lambda\lambda$1548,1550 emission from the central 1-arcsec 
spectrum. This is expected as spatial variation of the gas kinematics would broaden the overall 
line profile. The fitted 1$\sigma$ dispersion of spatially-integrated 
\ion{He}{II}\,$\lambda$1640 is consistent with that of the extracted narrow 
\ion{He}{II}\,$\lambda$1640 from the central 1-arcsec spectrum. This is expected as the detectable 
\ion{He}{II}\,$\lambda$1640 has a small spatial extent that is only marginally larger than the 
point spread function. 

\subsubsection{Spatially Resolved Kinematics} 

\begin{figure*}
\includegraphics[width=0.36\textwidth]{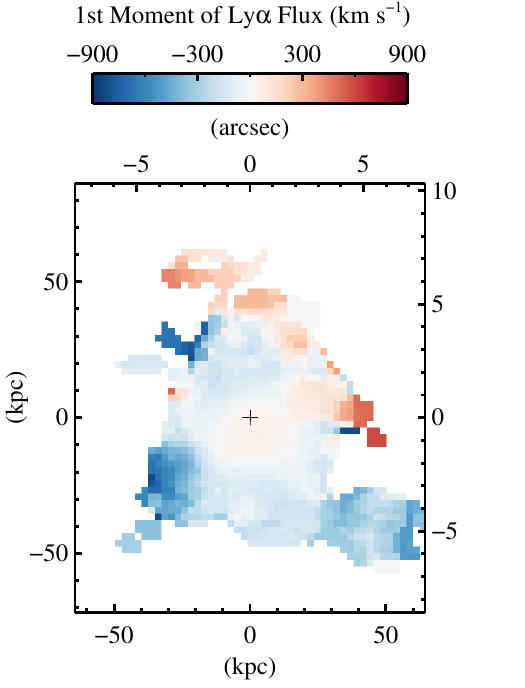}
\hspace{-0.3in}
\includegraphics[width=0.36\textwidth]{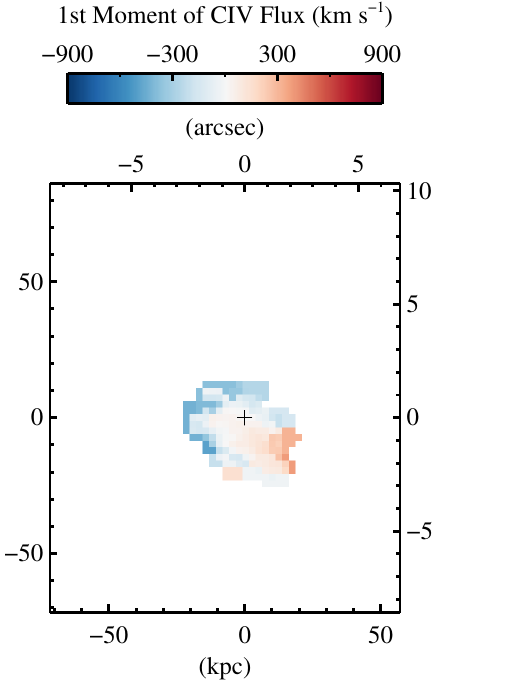}
\hspace{-0.4in}
\includegraphics[width=0.36\textwidth]{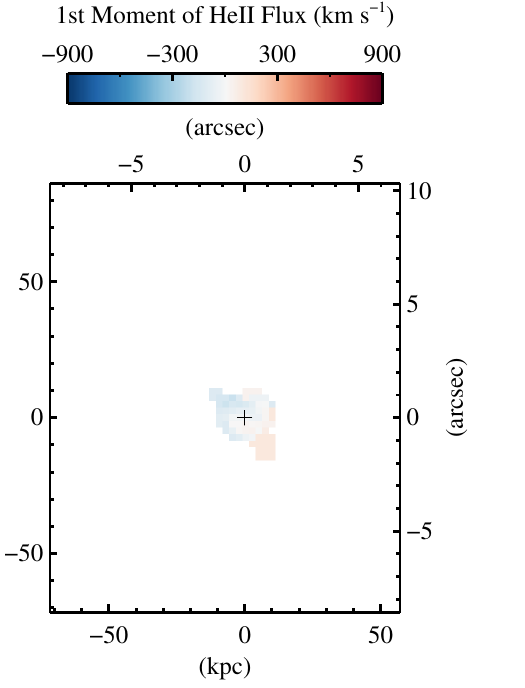}
\caption{Line velocity centroid maps of Ly$\alpha$, \ion{C}{IV}\,$\lambda$1549, and 
\ion{He}{II}\,$\lambda$1640, obtained as the first velocity moments of the line flux 
distributions. For each emission line, the values are calculated using only voxels that are 
significantly detected, i.e.\ selected by the three-dimensional segmentation mask. 
}
\label{fig:1stmmnt}
\end{figure*}

We create velocity centroid maps obtained as the first moments in velocity space of the flux 
distribution at each spatial position. To produce these two-dimensional maps we only include 
significantly detected voxels selected by a line's three-dimensional segmentation mask.  
The maps are presented in Fig.~\ref{fig:1stmmnt}. Summary statistics of the maps are presented 
in Table~\ref{tab:resolvedkin}. 

From the velocity centroid maps, the majority of the line emissions have small shifts relative to 
the systemic velocity. With the exception of regions associated with low S/N emissions, the 
velocity shift of detectable \ion{C}{IV}\,$\lambda$1549 and \ion{He}{II}\,$\lambda$1640 are close 
to the velocity shift measured for Ly$\alpha$ in the same spatial locations, typically within one 
spectral resolution element. The angular size of the fitted point spread function of the quasar 
continuum plus broad-line region 1.4\,arcsec corresponds to a physical distance of 12\,kpc. The 
velocity fields are thus quiet on scales down to the innermost $\sim$12\,kpc in projected linear 
extent. As the extended Ly$\alpha$ emission has good S/N covering a large area, we may examine its 
velocity field further. The velocity field is coherent in the inner halo region on scales 
$\lesssim$(20\textendash30)\,kpc from the halo centroid or the quasar. The halo shows clear 
velocity shear from the northern edge to the southern edge. The gradient is of an order two times 
the estimated one-dimensional velocity dispersion of the dark matter halo. Rotating disc of 
$\sim$100\,kpc size is not expected to be in place by $z\sim2$ 
\citep[e.g.,][]{DeFelippis+20,Huscher+21}. Together with the filamentary morphology, the northern 
and southern edges are most consistent with tracing inflowing streams. 

\begin{figure}
\includegraphics[width=0.36\textwidth]{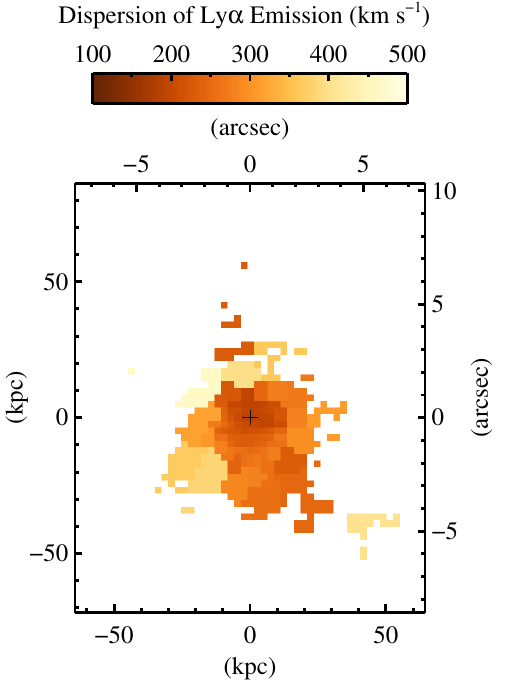}
\caption{Ly$\alpha$ velocity dispersion map, obtained by fitting Gaussians on Voronoi-binned 
data. 
}
\label{fig:vordisp}
\end{figure}

There is not enough signal in the halo to create maps of the second moments in velocity 
space of the flux distribution \citep[see discussions in][]{OSullivan+20}. If only the 
significantly detected voxels are included in calculating the second moments, wings of the line 
profiles cannot be captured and the second moment values would be biased low. If all voxels are 
included, the second moment values would be dominated by noise at large relative velocities. We 
therefore measure the Ly$\alpha$ velocity dispersion map by fitting Gaussians to a Voronoi-binned 
map \citep[see][]{Rupke+19}. From Ly$\alpha$ line flux and error maps we construct Voronoi bins 
with a target S/N of 10 and a threshold S/N of 1. We then fit single Gaussians to the 
Voronoi-binned data. The Ly$\alpha$ Gaussian velocity dispersion map is presented in 
Fig.~\ref{fig:vordisp} and the summary statistics are presented in Table~\ref{tab:resolvedkin}. 
For the extended \ion{C}{IV}\,$\lambda$1549 or \ion{He}{II}\,$\lambda$1640, there is not enough 
signal to spatially resolve their velocity dispersions. 

From the Ly$\alpha$ velocity dispersion map, the majority of the emission has velocity dispersion 
comparable to the estimated one-dimensional velocity dispersion of the dark matter halo, typically 
within 10 per cent. The halo shows some increase in velocity width from the inner region to the 
outer region. The change is of an order the estimated one-dimensional dark matter velocity 
dispersion. 
The velocity dispersion of the Ly$\alpha$ emission is coherent across distant spatial locations, 
and is consistent with gravitational motions. There is no clear evidence of gravitationally 
unbound outflows down to half the size of the point spread function $\sim$0.7\,arcsec or 
$\sim$6\,kpc scale from the quasar. 

\begin{figure}
\includegraphics[width=0.36\textwidth]{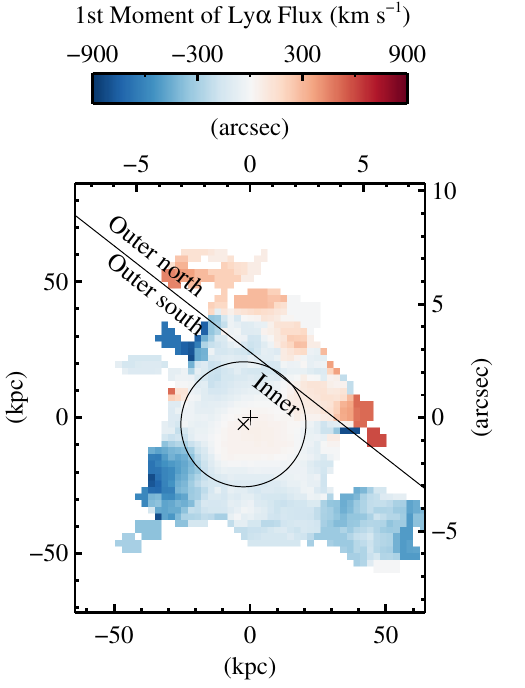}
\includegraphics[width=\columnwidth]{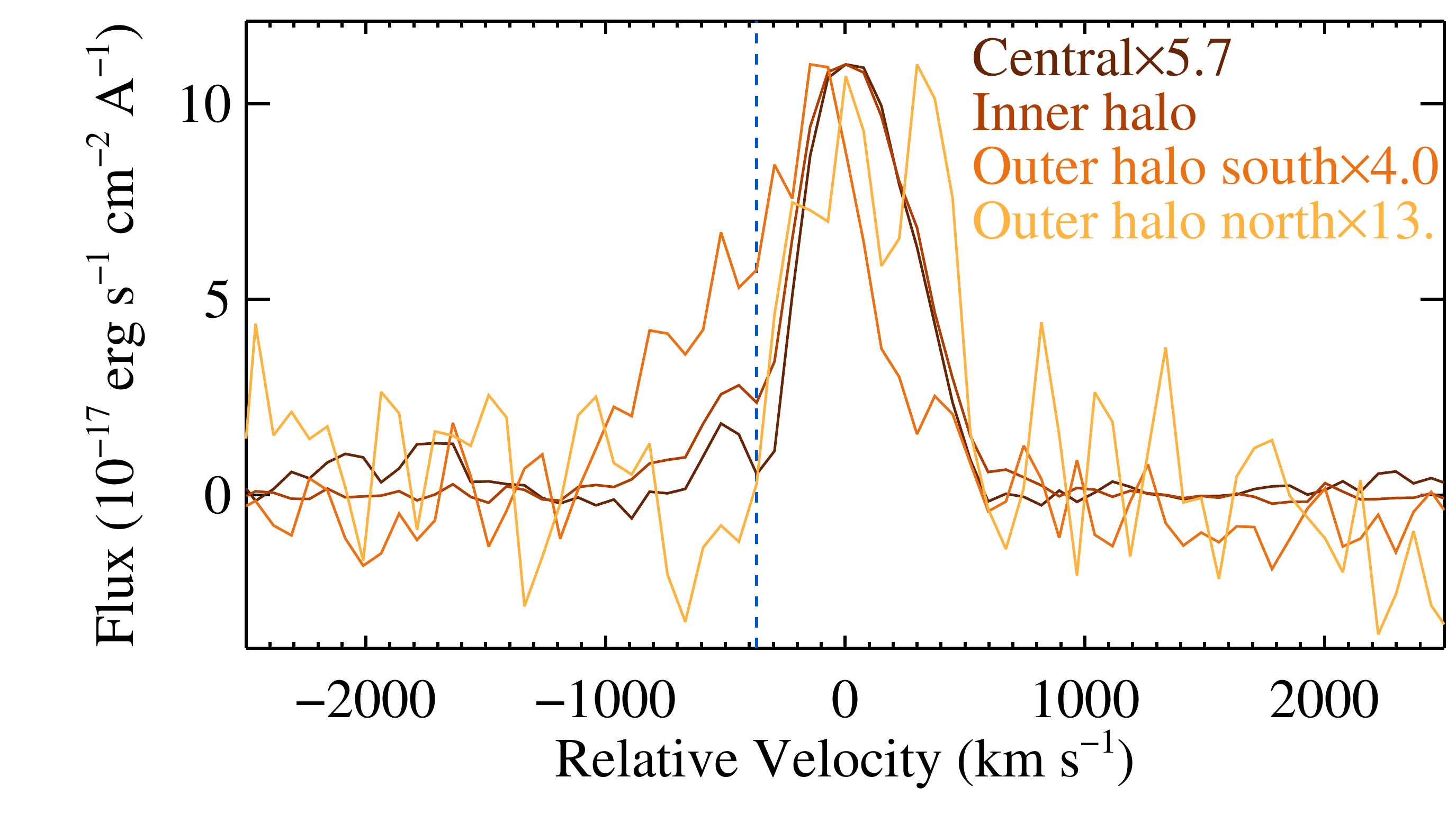}
\caption{One-dimensional spectra extracted from apertures defined using features on the first 
velocity moment map of Ly$\alpha$. The top panel shows the delineations on the first 
moment map. The plus symbol marks the quasar position. The cross symbol marks the halo peak.  
The circle marks an inner halo region of radius 23\,kpc extending from the halo peak, where the 
line-emitting gas is largely at rest relative to the systemic velocity. For the outer halo, a 
slanted line divides it into a southern region and a northern region along the change from 
negative to positive velocities. The bottom panel shows the spectra extracted from these three 
apertures, and we overplot the central 1-arcsec aperture spectrum in the same panel. To facilitate 
comparison of the profiles, the central 1-arcsec spectrum, the southern outer halo spectrum, and 
the northern outer halo spectrum are exaggerated. The blue dashed line marks the relative velocity 
of the absorption trough in the inner halo spectrum.}
\label{fig:aperspecs}
\end{figure}

Given the velocity shear in the outer Ly$\alpha$ halo and the coherence and quietness in the inner 
Ly$\alpha$ halo, we define three apertures according to the features on the velocity centroid map 
and the morphology to further examine the kinematics, as shown in Fig.~\ref{fig:aperspecs}. We 
delineate an outer halo region and an inner halo region with a circle of radius 23\,kpc centred on 
the halo's peak. We further divide the outer halo into a southern region and a northern region 
with a slanted line that follows the transition from negative to positive relative velocities. In 
Fig.~\ref{fig:aperspecs} we present the one-dimensional spectra extracted from these apertures, 
and we overplot the central 1-arcsec aperture spectrum on this figure. We find that the Ly$\alpha$ 
flux in the central 1\,arcsec does not dominate the total Ly$\alpha$ flux in the defined inner halo 
aperture. The Ly$\alpha$ spike in the central 1-arcsec spectrum has a line profile very similar to 
the Ly$\alpha$ emission in the inner halo aperture, confirming that the spike originates from the 
inner halo region. The Ly$\alpha$ line profiles of the central 1-arcsec and inner halo spectra, 
which carry the same blueshifted absorption feature, resemble those of carefully modeled, 
spatially resolved Ly$\alpha$ halo spectra in the literature that require the presence of outflows 
\citep[e.g.,][]{Li+21}. The blueshifted absorption feature is absent in the two outer halo 
spectra. All three halo aperture spectra show evidence of multiple kinematic components, 
suggesting that the emitting gas is clumpy. 

We identify a narrow absorption feature in the central 1-arcsec spectrum, whose trough relative to 
the systemic has a velocity of $-$369\,km\,s$^{-1}$. This absorption feature is spatially coherent 
across an extent matching the defined inner halo. Assuming that the blueshifted absorption traces 
a simple isotropic expanding shell-like structure, this value reflects a characteristic outflow 
speed. 
While J0006+1215 displays [\ion{O}{III}]\,$\lambda$5007 at speeds and velocity widths 
$\sim$6000\,km\,s$^{-1}$ emitting at $\sim$1\,kpc, there is no evidence for this fast flow on halo 
scales as probed by the KCWI data. For outflows at the measured moderate speeds that are present 
on scales out to (20\textendash30)\,kpc, the dynamical time involved will be 
$\sim$(5\textendash8)$\times10^7$ yr. 

\begin{table}
\caption{Summary of Spatially Resolved Velocity Centroids and Dispersions of the Extended Emission}
\label{tab:resolvedkin}
\setlength{\tabcolsep}{0.1in}
\begin{tabular}{lcc}
\hline
Line Name & Spatial Median & Spatial Standard Deviation \\
 & (km\,s$^{-1}$) & (km\,s$^{-1}$) \\
\hline
\multicolumn{1}{l}{} & \multicolumn{2}{c}{Velocity Centroid Map} \\
\hline
\ion{H}{I}\,Ly$\alpha$ & $-85$ & $235$ \\
\ion{C}{IV}\,$\lambda$1549 & $-23$ & $201$ \\
\ion{He}{II}\,$\lambda$1640 & $-9$ & $87$ \\
\hline
\multicolumn{1}{l}{} & \multicolumn{2}{c}{Fitted Gaussian Velocity Dispersion Map} \\
\hline
\ion{H}{I}\,Ly$\alpha$ & $287$ & $70$ \\
\hline
\end{tabular}
\end{table}

\subsection{Comparison to Other Quasars}

\subsubsection{Introducing Comparison Samples and Single Sources}

We compare the J0006+1215 line-emitting halo with other samples and single sources from the 
literature. They are line-emitting haloes surrounding quasars selected by various criteria and are 
crudely matched in bolometric luminosity and cosmic redshift to J0006+1215, i.e.\ quasars of 
$\gtrsim$10$^{46}$\,erg\,s$^{-1}$ at cosmic noon. When available we calculate or adopt their halo 
redshifts, elliptical eccentricity parameters, maximum linear sizes, line luminosities, surface 
brightness radial profiles, inner halo line ratios, and spatially-integrated velocity dispersions. 
We estimate all intrinsic bolometric luminosities from {\it W}3-band photometries. 
We apply Galactic extinction corrections to all line luminosities and surface brightnesses. 
In Table~\ref{tab:lit} we list the quantities we adopt from these literature samples and single 
sources. Below are our remarks about them. 

The \cite{Cai+19} sample consists of 16 luminous blue quasars of median $z=2.3$ and median
bolometric luminosity 10$^{47.2}$\,erg\,s$^{-1}$.
We adopt their sample median Ly$\alpha$ surface brightness radial profile.
The $e_{\rm weight}$ values are calculated with a 1-arcsec region centred on the quasar masked
out to avoid residuals from point-spread function subtraction.

The \cite{ArrigoniBattaia+19} sample consists of 61 luminous blue quasars of median $z=3.2$
and median bolometric luminosity 10$^{47.4}$\,erg\,s$^{-1}$.
For each halo we estimate its maximum linear size as the sum of maximal radial extent from
quasar plus $\sqrt{{\rm (covered\,area)}/{\rm \pi}}$.
We adopt their sample average Ly$\alpha$ surface brightness radial profile.
The $e_{\rm weight}$ values are calculated with the central 1-arcsec region masked.

The \cite{Borisova+16a} sample consists of 19 luminous blue quasars of median $z=3.2$ and
median bolometric luminosity 10$^{47.6}$\,erg\,s$^{-1}$.
We adopt their sample median Ly$\alpha$ surface brightness radial profile as obtained from
\cite{Marino+19}. The $e_{\rm unweight}$ values of this sample are obtained from
\cite{Mackenzie+21}. They are calculated with a 1-arcsec region centred on the quasar masked.

The \cite{Guo+20} sample consists of 80 luminous blue quasars of median $z=3.2$ and median
bolometric luminosity 10$^{47.5}$\,erg\,s$^{-1}$. It is the combined sample of
\cite{ArrigoniBattaia+19} and \cite{Borisova+16a}. We adopt their sample stacked
\ion{C}{IV}\,$\lambda$1549 surface brightness radial profile. We adopt their sample
stacked \ion{C}{IV}\,$\lambda$1549/Ly$\alpha$ and \ion{He}{II}\,$\lambda$1640/Ly$\alpha$ ratios of
an inner region $\sim$20\,kpc in size.

The \cite{Mackenzie+21} sample consists of 12 fainter blue quasars of median $z=3.2$ and
median bolometric luminosity 10$^{47.0}$\,erg\,s$^{-1}$.
We adopt their sample average Ly$\alpha$ surface brightness radial profile.
The $e_{\rm unweight}$ values are calculated with the central 1-arcsec region masked.

The \cite{denBrok+20} sample consists of four radio-quiet Type~II active galactic nuclei of median
$z=3.4$ and median intrinsic bolometric luminosity 10$^{46.7}$\,erg\,s$^{-1}$. The sources
are X-ray selected. We verify that the intrinsic bolometric luminosities estimated based on
measured mid-IR luminosities are similar to those estimated based on measured X-ray luminosities
and the correction factors modeled in \cite{Shen+20}.
We measure the maximum linear sizes of the Ly$\alpha$ haloes by hand on the maps. Two of their
sources have significantly longer exposure times than J0006+1215 and their sizes are omitted in
the comparison.
We adopt the individual Ly$\alpha$ surface brightness radial profiles of the sources. As their
sources are obscured, no point-spread function subtraction is needed to reveal halo emissions.
\ion{C}{IV}\,$\lambda$1549/Ly$\alpha$ ratios are not reported and we plot them as lower limits
above zero. For two of their sources which have \ion{He}{II}\,$\lambda$1640 maps presented, we
measure the \ion{He}{II}\,$\lambda$1640/Ly$\alpha$ ratios by hand in a region $\sim$10\,kpc in
size around the sources. Their $e_{\rm unweight}$ values are calculated without masking the
centre.

The single source of \cite{Marino+19} is a luminous blue quasar absorbed by a proximate damped
Ly$\alpha$ system at $z=3.0$.
We adopt the \ion{C}{IV}\,$\lambda$1549/Ly$\alpha$ and \ion{He}{II}\,$\lambda$1640/Ly$\alpha$
ratios of a region $\sim$20\,kpc in size around the central source.


The single source of \cite{Sanderson+21} is a radio-quiet Type~II active galactic nucleus at
$z=3.2$. The Ly$\alpha$ halo was discovered before the mid-IR central source was
identified. The mid-IR source is not detected in the optical. We adopt the
\ion{C}{IV}\,$\lambda$1549/Ly$\alpha$ and \ion{He}{II}\,$\lambda$1640/Ly$\alpha$ ratios of a
region $\sim$20\,kpc in size around the central source. Thhe $e_{\rm unweight}$ value is
calculated without masking the centre.


The single source of \cite{Cai+17} is a radio-quiet Type~II active galactic nucleus at $z=2.3$. We
adopt the \ion{C}{IV}\,$\lambda$1549/Ly$\alpha$ and \ion{He}{II}\,$\lambda$1640/Ly$\alpha$ ratios
of a region $\sim$10\,kpc in size that is offset by 15\,kpc from the source.


In general, measurements that describe an entire halo are not sensitive to masking the central
1\,arcsec of the Ly$\alpha$ halo, which is a necessary procedure for studying haloes powered by
blue quasars. It is fair to compare quantities that describe the overall halo of J0006+1215 and 
haloes in the literature. In J0006+1215, the central source is obscured, allowing views of the 
innermost halo region which is not possible with blue quasars. As demonstrated in 
\cite{Mackenzie+21}, even for fainter blue quasars, uncertain point-spread function subtraction 
can dilute physical trends between quasar properties and halo properties on scales 
$\lesssim$20\,kpc from the quasar. Comparing quantities that are sensitive to the inner halo 
requires caution. 

\subsubsection{The Comparison Results}

\begin{figure}
\includegraphics[width=\columnwidth]{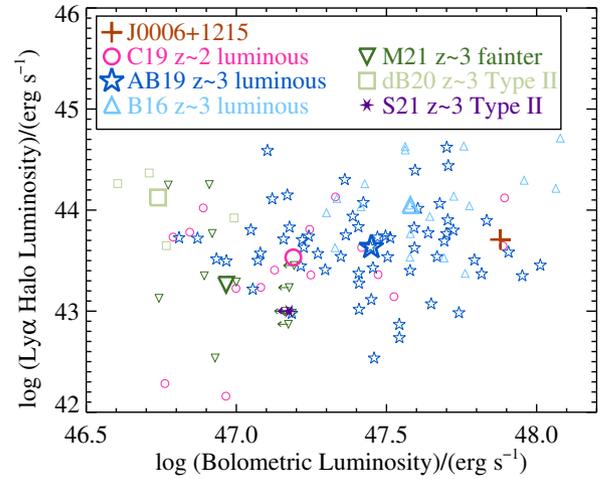}
\caption{Ly$\alpha$ halo luminosity against quasar bolometric luminosity, of various literature 
sources and J0006+1215. Upper limits and lower limits on the quantities are shown with arrows. The 
brown plus symbol shows J0006+1215. The dark pink circle symbols, the dark blue star symbols, the 
light blue upward triangle symbols, and the dark green downward triangle symbols show samples of 
Type~I blue quasars. The light green square symbols show a sample of Type~II active galactic 
nuclei. Medians of these literature samples are plotted as larger symbols of the same styles as 
their individual data points. The dark purple hexagram symbol shows a single Type~II source.}
\label{fig:neblum_Lbol}
\end{figure}

Fig.~\ref{fig:neblum_Lbol} compares J0006+1215's Ly$\alpha$ halo luminosity with sources from the 
literature. We have verified that the calculated Ly$\alpha$ halo  
luminosities of J0006+1215 without and with masking the central 1\,arcsec are very similar, and 
the central 1-arcsec line luminosity is only 10 per cent of the total. Thus we may compare 
Ly$\alpha$ halo luminosities of different sources despite that the innermost halo emissions 
of some of them cannot be recovered. 
Considering only the Type~I blue quasars in the figure, halo line luminosity weakly increases with 
quasar bolometric luminosity with substantial scatter in the trend. If one fits a linear relation 
between the halo luminosities and bolometric luminosities of the Type~I blue quasars, the measured 
halo luminosity of J0006+1215 would be $\sim$0.3 times the prediction from this relation. This 
factor of three deficit in halo luminosity for J0006+1215 is well within the scatter around the 
trend between halo luminosity and bolometric luminosity. 

\begin{figure}
\includegraphics[width=\columnwidth]{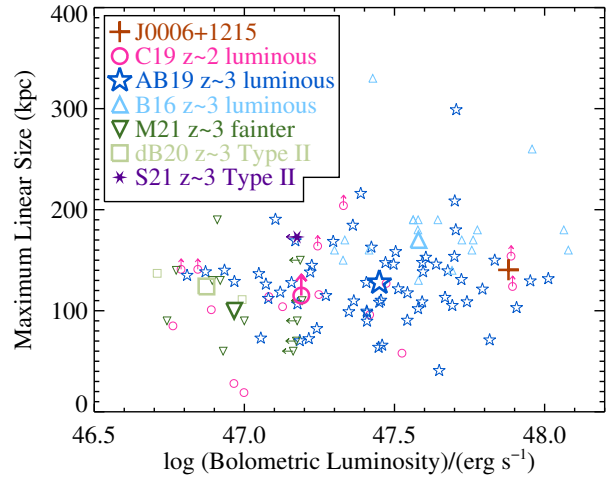}
\caption{Maximum linear size of Ly$\alpha$ halo against quasar bolometric luminosity of various 
literature sources and J0006+1215. The symboling scheme follows Fig.~\ref{fig:neblum_Lbol}.}
\label{fig:linsz_Lbol}
\end{figure}

Fig.~\ref{fig:linsz_Lbol} compares J0006+1215's halo maximum linear size with sources from the 
literature. The J0006+1215 halo's extent is not outstanding. 

\begin{figure}
\includegraphics[width=\columnwidth]{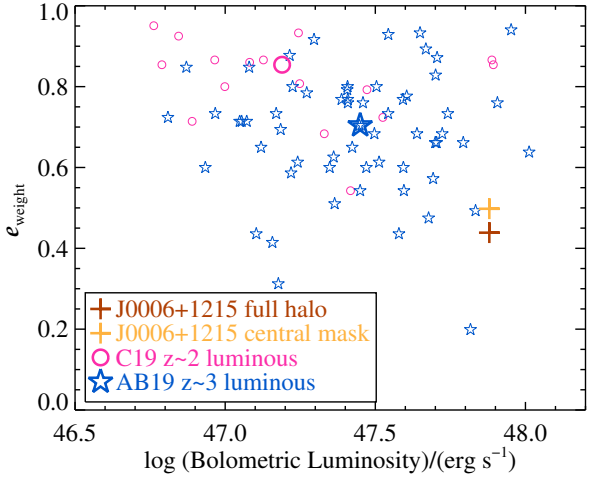}
\caption{Flux-weighted elliptical eccentricity against quasar bolometric luminosity, of various 
literature sources and J0006+1215. The symboling scheme follows Fig.~\ref{fig:neblum_Lbol}. For 
J0006+1215, we show both the $e_{\rm weight}$ value calculated with the full Ly$\alpha$ halo 
and that calculated with the central 1\,arcsec masked for fairer comparison with blue quasars.}
\label{fig:ewght_Lbol}
\end{figure}

Fig.~\ref{fig:ewght_Lbol} compares J0006+1215's flux-weighted elliptical eccentricity with sources 
from the literature. Since $e_{\rm weight}$ is sensitive to inner halo emission, for 
fairer comparison with blue quasars we show both the J0006+1215 $e_{\rm weight}$ value calculated 
with the full Ly$\alpha$ halo, 0.44, and that calculated with the central 1\,arcsec masked, 0.50.  
Both calculations yield a low value relative to other sources. The J0006+1215 inner halo is more 
circularly symmetric than the inner haloes of most luminous blue quasars. 

\begin{figure}
\includegraphics[width=\columnwidth]{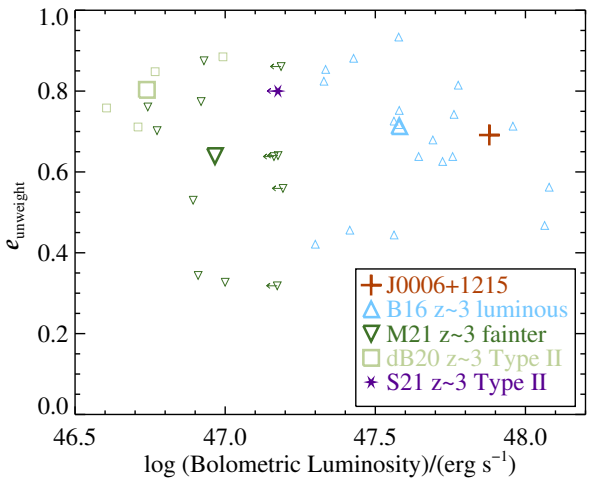}
\caption{Flux-unweighted elliptical eccentricity against quasar bolometric luminosity, of various 
literature sources and J0006+1215. The symboling scheme follows Fig.~\ref{fig:neblum_Lbol}.}
\label{fig:eunwght_Lbol}
\end{figure}

Fig.~\ref{fig:eunwght_Lbol} compares J0006+1215's flux-unweighted elliptical eccentricity with 
sources from the literature. Our comparison holds despite that 
the innermost halo emissions of some of the literature sources cannot be recovered, since the 
overall halo eccentricity is insensitive to the innermost emission. For J0006+1215, without or 
with masking the central 1\,arcsec yields nearly the same $e_{\rm unweight}$. 
The J0006+1215 halo's overall morphology is similarly asymmetric to the average of haloes 
surrounding luminous blue quasars and fainter blue quasars, and is less asymmetric than the average 
of haloes surrounding Type~II active galactic nuclei. 

\begin{figure}
\includegraphics[width=\columnwidth]{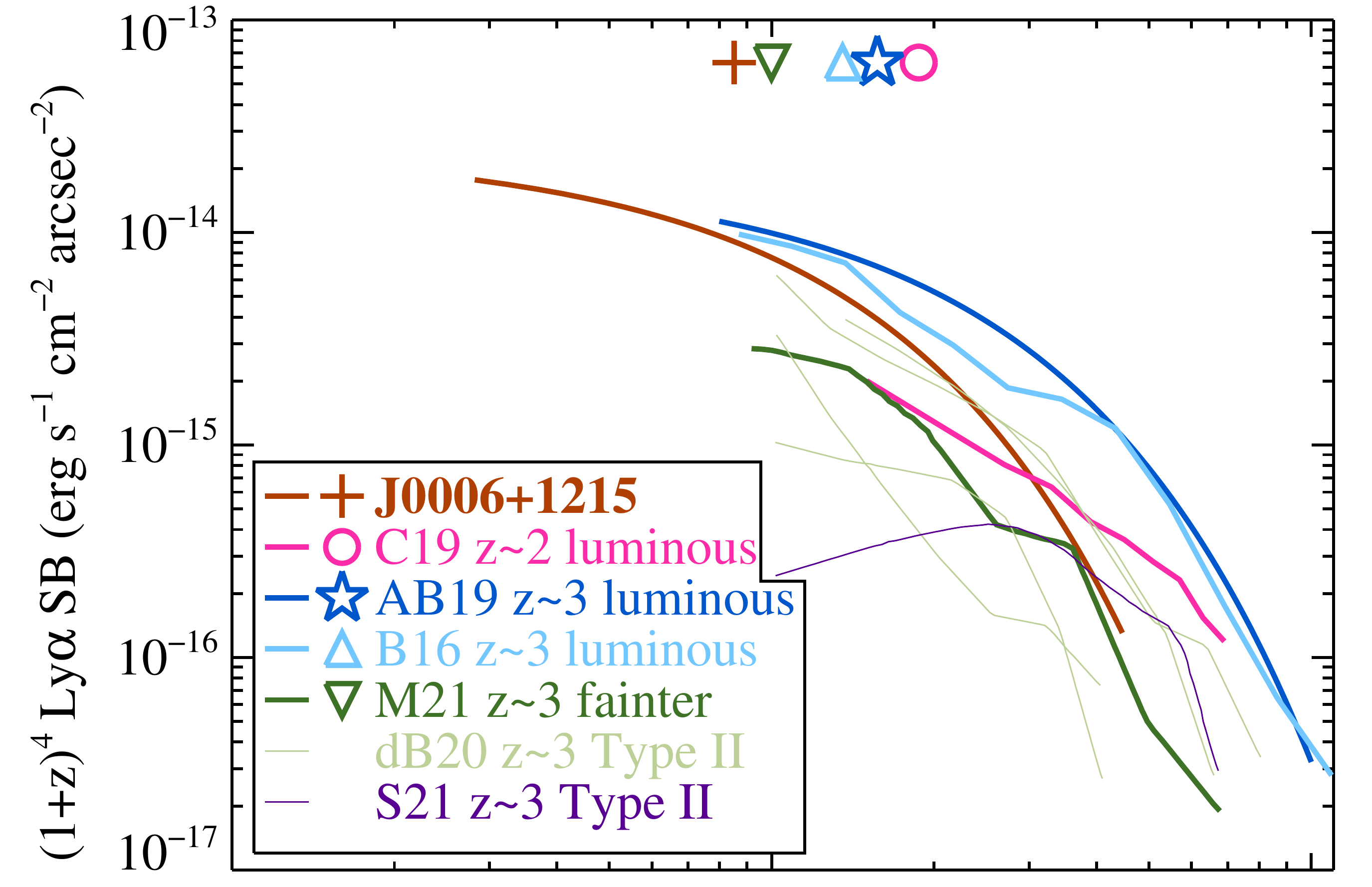}
\includegraphics[width=\columnwidth]{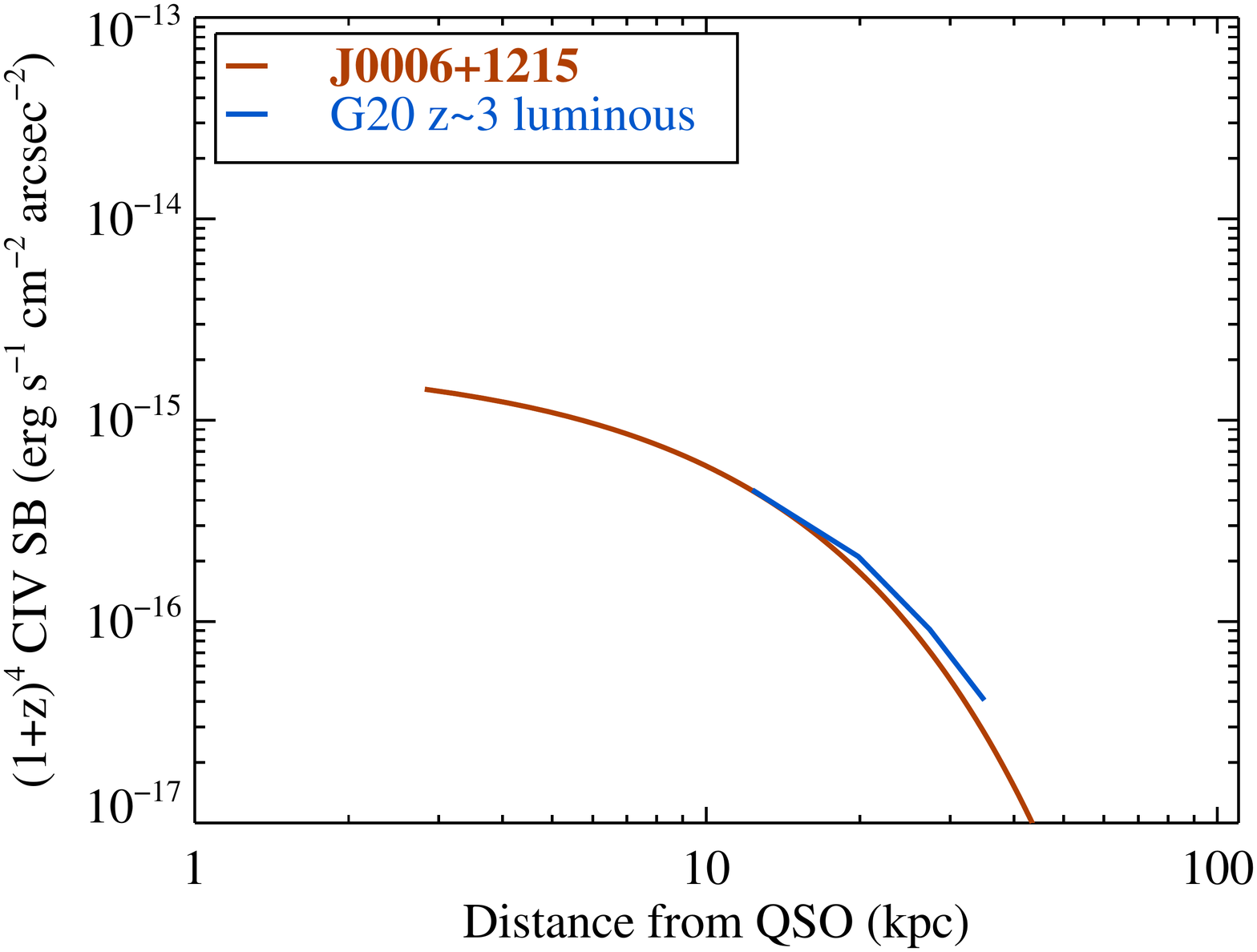}
\caption{Surface brightness radial profiles of Ly$\alpha$ and \ion{C}{IV}\,$\lambda$1549 of 
various literature sources and the exponential fits to the surface brightness radial profiles of 
Ly$\alpha$ and \ion{C}{IV}\,$\lambda$1549 of J0006+1215. Top panel shows Ly$\alpha$. The 
brown curve shows J0006+1215. The dark pink curve, the dark blue curve, the light blue curve, and 
the dark green curve are averages of samples of Type~I blue quasars. The thin light green curves 
are four different sources from a sample of Type~II active galactic nuclei. The thin dark purple 
curve is a single Type~II source. Near the top edge of this panel we plot the exponential scale 
lengths of the blue quasar samples and J0006+1215. 
Bottom panel shows \ion{C}{IV}\,$\lambda$1549. The brown curve shows J0006+1215. The dark 
blue curve is the stack a sample of luminous blue quasars. Axes of the bottom panel displays the 
same data ranges as the top panel. displays All surface brightnesses are corrected for 
cosmological dimming.}
\label{fig:SBradial_lit}
\end{figure}

Fig.~\ref{fig:SBradial_lit} compares the fit to J0006+1215's Ly$\alpha$ surface brightness radial 
profile and the fit to J0006+1215's \ion{C}{IV}\,$\lambda$1549 surface brightness radial profile 
with sources from the literature. 
The obscuration in J0006+1215 allows tracing the innermost region of the halo, which 
is not possible for normal blue quasars whether faint or luminous. 
The exponential function fit to the J0006+1215 Ly$\alpha$ halo profile has a relatively small 
scale length of 8.5\,kpc. In contrast the comparison blue quasar samples are described by 
exponential scale lengths of 10.0\,kpc to 18.7\,kpc. Relative to blue quasars the J0006+1215 
Ly$\alpha$ halo is more centrally concentrated. We note that the smaller scale length of 
J0006+1215 is not caused by recovering innermost halo emission. Half the FWHM size of the 
point-spread function in our KCWI data is 0.7\,arcsec or 6\,kpc, and the comparison data have 
similar point-spread function sizes. Visual comparison of the surface brightness radial profiles 
beyond 6\,kpc projected distance reveals that the J0006+1215 Ly$\alpha$ profile is indeed steeper. 
The J0006+1215 \ion{C}{IV}\,$\lambda$1549 profile resembles that of the luminous blue quasars. 

\begin{figure}
\includegraphics[width=\columnwidth]{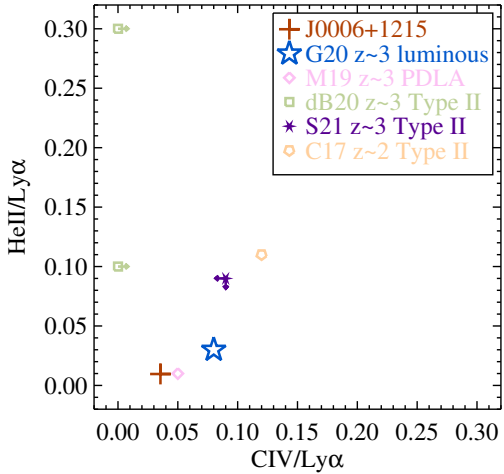}
\caption{\ion{He}{II}\,$\lambda$1640/Ly$\alpha$ ratio against 
\ion{C}{IV}\,$\lambda$1549/Ly$\alpha$ ratio spatially integrated over inner halo regions where 
both lines are significantly detected, of various literature sources and J0006+1215. The brown 
plus symbol shows J0006+1215. The large dark blue star symbol shows the stack of a sample of 
Type~I blue quasars. The light pink diamond symbol shows a single Type~I blue quasar. The light 
green square symbols, the dark purple hexagram symbol, and light orange downard pentagon symbol 
show Type~II sources from different studies. Both data axes show the same range to keep a 1:1 
aspect ratio.}
\label{fig:HeIILya_CIVLya}
\end{figure}

Fig.~\ref{fig:HeIILya_CIVLya} compares J0006+1215's \ion{C}{IV}\,$\lambda$1549/Ly$\alpha$ and 
\ion{He}{II}\,$\lambda$1640/Ly$\alpha$ ratios with sources from the literature. 
The ratios are reported for line fluxes measured over inner regions $\sim$(10\textendash20)\,kpc 
in sizes. Compared to the inner haloes around other Type~I quasars which have 
\ion{He}{II}\,$\lambda$1640/Ly$\alpha$ ratios $\sim$(0.01\textendash0.03),  
the J0006+1215 inner halo has a similar line ratio. Compared to the 
inner haloes of radio-quiet Type~II active galactic nuclei which have 
\ion{He}{II}\,$\lambda$1640/Ly$\alpha$ ratios reaching $\sim$0.1,  
the inner haloes of J0006+1215 and other Type~I quasars have lower line ratios. 

\begin{figure}
\includegraphics[width=\columnwidth]{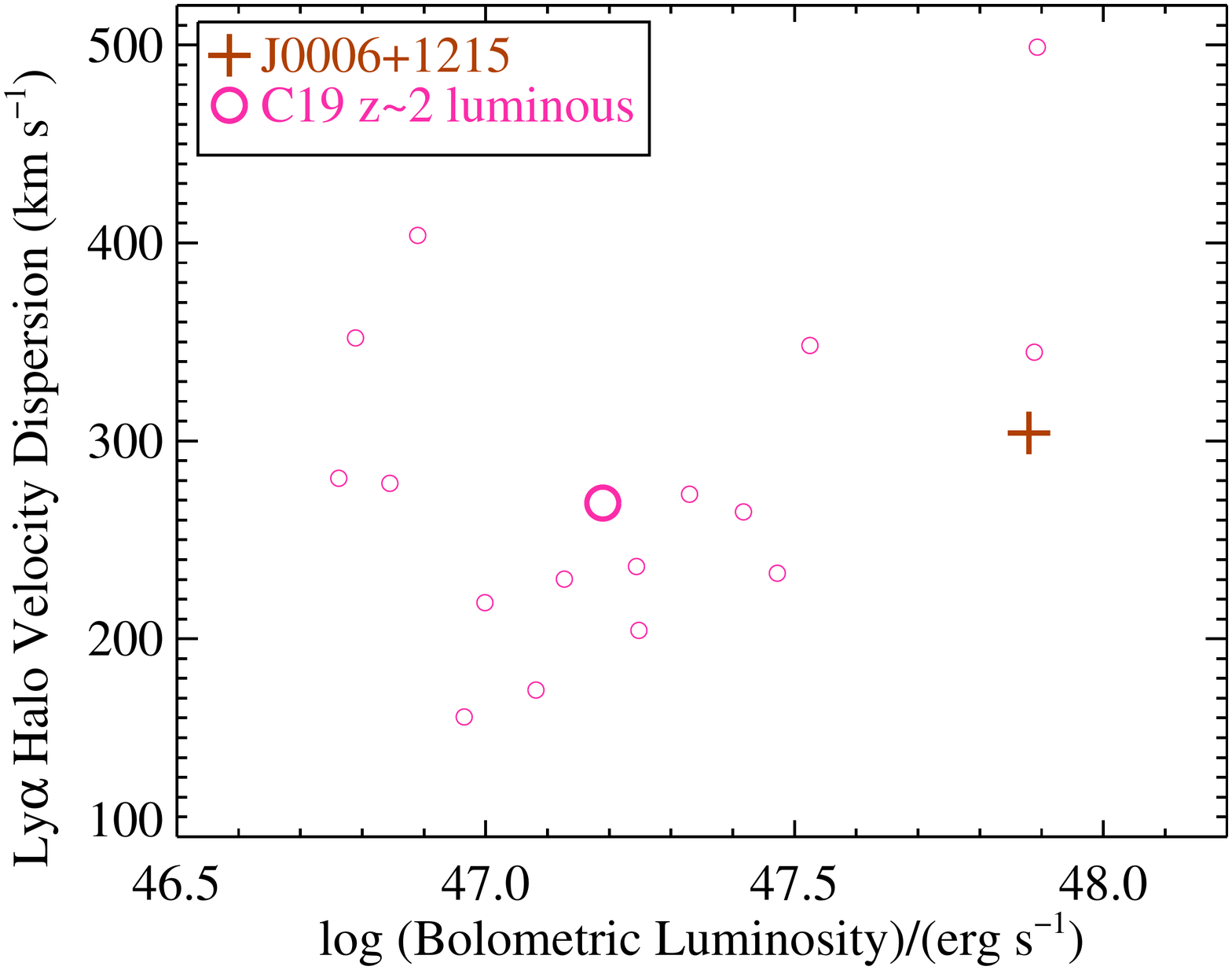}
\caption{Spatially integrated velocity dispersion of the Ly$\alpha$ halo emission against quasar 
bolometric luminosity, of literature sources and J0006+1215. The brown symbol shows J0006+1215. 
The dark pink symbols show a sample of luminous blue quasars, whose median is plotted as a larger 
symbol of the same style.}
\label{fig:disp_Lbol}
\end{figure}

Fig.~\ref{fig:disp_Lbol} compares J0006+1215's Ly$\alpha$ halo-integrated velocity dispersion 
with sources from the literature.  
With typical $\sigma_{\rm 1D}$ values of quasar host dark matter haloes as estimated in Section 
3.4 as baseline comparison, the J0006+1215 halo is similarly kinematically quiet to the average of 
haloes of luminous blue quasars at similar redshifts. 

\begin{table*}
\caption{Literature Samples and Single Sources for Comparison}
\label{tab:lit}
\setlength{\tabcolsep}{0.01in}
\begin{tabular}{lcccc}
\hline
Reference & Sample Size and Source Type & Data Type & Adopted Measurements \\
\hline
\citet{Cai+19} & 16 luminous blue quasars & Integral field & Halo luminosity, size, $e_{\rm weight}$, average SB profile, velocity dispersion \\
\citet{ArrigoniBattaia+19} & 61 luminous blue quasars & Integral field & Halo luminosity, size, $e_{\rm weight}$, average SB profile \\
\citet{Borisova+16a} & 19 luminous blue quasars & Integral field & Halo luminosity, size, $e_{\rm unweight}$, average SB profile \\
\citet{Guo+20} & 80 luminous blue quasars & Integral field & \ion{C}{IV} average SB profile, averageline ratios \\
\citet{Mackenzie+21} & 12 fainter blue quasars & Integral field & Halo luminosity, size, $e_{\rm unweight}$, average SB profile \\ 
\citet{denBrok+20} & 4 Type II AGN & Integral field & Halo luminosity, size, $e_{\rm unweight}$, individual SB profiles, individual line ratios \\
\citet{Marino+19} & Single proximate DLA quasar & Integral field &  Line ratios \\
\citet{Sanderson+21} & Single Type II AGN & Integral field & Halo luminosity, size, $e_{\rm unweight}$, SB profile, line ratios \\ 
\citet{Cai+17} & Single Type II AGN & Narrowband image & Line ratios \\
\hline
\end{tabular}
\end{table*}

\section{Discussion} 

\subsection{Measuring Quasar Systemic Redshifts}

With the aid of dust obscuration in J0006+1215 to reveal narrow line components, in Section 3.4 we
show that the redshift and line profile of the narrow component in Ly$\alpha$ in the central
1\,arcsec match the inner halo emission. Therefore this spike also arises from the inner halo
region, with the central 1-arcsec emission otherwise dominated by quasar emission.
Redshifts and profile widths of the narrow component in the central 1-arcsec Ly$\alpha$ and that
in the unresolved [\ion{O}{III}]\,$\lambda$5007 are both consistent within one spectral resolution
element $\approx$150\,km\,s$^{-1}$. This confirms that narrow [\ion{O}{III}]\,$\lambda$5007
component measured in the central 1-arcsec aperture is also related to the inner halo.
These narrow emission lines arising from the kinematically quiet inner halo region should be good
indicators of the systemic redshift, and therefore good velocity references for measuring 
blueshifts in the quasar broad emission lines. This is particularly helpful for an ERQ, 
as blueshifts are larger and more pervasive across different rest-frame UV lines than 
normal blue quasars. In J0006+1215, the broad \ion{C}{IV}\,$\lambda$1549's velocity centroid is
blueshifted by 2240\,km\,s$^{-1}$.

If the expectation of no large velocity shifts between the haloes and the systemics generally
applies, then narrow features in quasar spectra can accurately reveal large blueshifts in the broad 
emission lines.
Having access to more reliable estimates of systemic redshifts also allows studying quasar-powered
haloes with the narrowband imaging technique, which had missed quasar-powered haloes in previous
studies \citep[e.g.,][]{ArrigoniBattaia+16}. For intermediate to high redshift quasars, whose
spectral data are often limited to rest-frame UV wavelengths, their systemic redshifts are often
estimated based on broad emission lines and are uncertain by hundreds of km\,s$^{-1}$. This
explains why literature samples often find large differences between the Ly$\alpha$ halo redshifts
and the estimated systemic redshifts. For fainter blue quasars, \cite{Mackenzie+21} find that the 
velocity of the peak of a quasar's Ly$\alpha$ line traces the velocity centroid of the Ly$\alpha$ 
halo better than any other estimates of the quasar systemic. Our finding that a Ly$\alpha$ spike i
n the ERQ spectrum is inner halo emission is consistent with their speculation that the narrow 
peak of a fainter blue quasar's Ly$\alpha$ line is due to contribution from the halo.

\subsection{Measuring the Inner Circumgalactic Medium} 

Enabled by the obscuration in J0006+1215, Figs~\ref{fig:corespecs} and~\ref{fig:fitcorespec} show 
that the narrow \ion{H}{I}\,Ly$\alpha$ spike arising from the inner halo is kinematically distinct 
from the quasar's broad Ly$\alpha$ emission line. We are able to study surface brightnesses, 
morphologies, and spatially resolved kinematics of three halo emission lines down to zero 
projected distance from the quasar position. 

In Sections 3.2 and 3.4 we show there are no abrupt changes in the innermost regions of the 
Ly$\alpha$ surface brightness map and velocity maps. This suggests a smooth transition between 
the interstellar medium and the inner circumgalactic medium, consistent with the finding in 
\cite{Marino+19} where a proximate damped Ly$\alpha$ system absorbs the glare of the quasar. The 
emission from their proximate damped Ly$\alpha$ system dominates the inner halo region, however. 
We contend that the halo emission surrounding J0006+1215 is more representative of the more 
general population of quasar-powered haloes.  

From the line ratio maps in Section 3.3 we find that both the 
\ion{C}{IV}\,$\lambda$1549/Ly$\alpha$ and \ion{He}{II}\,$\lambda$1640/Ly$\alpha$ ratios are higher 
near the quasar than further away. Some measurements on line-emitting haloes in the literature 
show lower \ion{C}{IV}\,$\lambda$1549/Ly$\alpha$ or \ion{He}{II}\,$\lambda$1640/Ly$\alpha$ at 
small distances of (1\textendash2)\,arcsec from their quasars than further away 
\citep[e.g.,][]{Marino+19,Travascio+20}. It is possible that for blue quasars, extraction of 
narrow \ion{C}{IV}\,$\lambda$1549 or \ion{He}{II}\,$\lambda$1640 is more prone to oversubtracting 
the unresolved quasar than extraction of narrow Ly$\alpha$, due to their fainter nature. We 
caution that interpretations of spatial gradients of line ratios in inner halo regions  
need to take into account of uncertainties from oversubtraction. 

We show that the halo emission properties transition near a distance of 
$\sim$(20\textendash30)\,kpc from the centre. The morphology analysis in Section 3.2 
shows that the halo emission is more circularly symmetric in the inner region and appears 
filamentary in the outer region. The line ratio analysis in Section 3.3 shows that metal 
enrichment is present in the inner region. The kinematics analysis in Section 3.4 shows that, the 
velocity field is more coherent and more quiet in the inner halo, and shows more shear and 
higher dispersion in the outer halo. The kinematics analysis further shows that, a blueshifted 
absorption feature is present in the inner-halo Ly$\alpha$ line profile, and is absent in the 
outer-halo line profile. 

Evidence of a transition in gaseous properties from inner halo to outer halo regions has been 
reported in the literature for other samples. \cite{Chen+20} use projected galaxy-galaxy pairs to 
study gaseous haloes of $z\sim2$ star forming galaxies in absorption. 
They find that outflows at radial speeds $\sim$600\,km\,s$^{-1}$ dominate the absorption signals 
at projected distances within $\sim$50\,kpc and inflows dominate at larger projected distances. 
The limitation with background sightline absorption spectroscopy is that 
sensitivity to the innermost halo region is lacking. \cite{Chen+21} observe the gaseous haloes of 
$z\sim2$ star forming galaxies in emission with KCWI. 
They find that outflows of radial speeds at several hundred km\,s$^{-1}$ dominate on distance 
scales out to $\sim$30\,kpc from the central galaxies. 
Their extended emissions are predominantly only powered by scattering of centrally generated 
photons, 
thus it is very challenging to detect any transition in halo emission properties on scales 
$\sim$30\,kpc and beyond around inactive galaxies. \cite{Li+21} study a single bright Ly$\alpha$ 
halo at $z\sim3$ suspectedly powered by a central source. 
They find that outflows at radial speeds $\sim$600\,km\,s$^{-1}$ dominating within $\sim$30\,kpc 
from the centre that transition to inflows dominating distance scales beyond. The central powering 
source is not detected however, possibly due to dust extinction or contamination by a foreground 
source. \cite{Guo+20} stack integral-field spectroscopy data of $z\sim3$ luminous blue quasars, 
and find the average metallicity is about constant within $\sim$40\,kpc from the central quasars 
and declines sharply beyond that. The limitation with studying blue quasars is again the lack of 
sensitivity to the innermost halo region. 
While the estimated distance scales of inner versus outer halo may vary with the sensitivity of 
the data and the diverse quasar/galaxy environments, a picture of a two-component circumgalactic 
medium is emerging from the literature. This two-component nature of the circumgalactic medium, 
has also been observed around a radio galaxy \citep{Vernet+17}. The J0006+1215 results strengthen 
this picture of moderate-speed outflow dominating inner halo and inflow dominating outer halo. 

\subsection{No Prominent Dependence between Halo Emission and Quasar Colour}

In Section 3.5.2 we show that the size, line luminosity, and kinematics measured on the 
J0006+1215 \ion{H}{I}\,Ly$\alpha$ halo are within the wide, diverse ranges measured for Ly$\alpha$ 
haloes around blue quasars of matching luminosities and redshifts. These comparisons suggest that 
the quasar illumination pattern in the circumgalactic medium and the circumgalactic medium 
properties 
of J0006+1215 are broadly similar to those of luminous blue quasars, and are not prominently 
affected by its quasar colour. 

In Section 3.5.2 we do find that J0006+1215 has a more centrally concentrated halo and a more 
circularly symmetric inner halo, with implications that we discuss below. 

\subsection{Youth of the ERQ}

Fig.~\ref{fig:SBradial_lit} compares surface brightness radial profiles and shows that the 
J0006+1215 Ly$\alpha$ halo is more centrally concentrated. Fig.~\ref{fig:ewght_Lbol} compares 
flux-weighted elliptical eccentricities, which are more sensitive to inner halo morphologies, 
shows that the inner region of the J0006+1215 Ly$\alpha$ halo has a more symmetric appearance. The 
higher central concentration of the surface brightness and the more symmetric appearance of the 
inner halo are consistent with the ERQ being in an earlier stage of evolution than luminous blue 
quasars at similar redshifts. As evident from the \ion{C}{IV}\,$\lambda$1549 detection, outflow 
materials centrally driven by a previous quasar episode or star formation are present in the inner 
halo. If the ERQ is younger, there is less time for outflow materials to expand to outer halo 
regions. The inner halo is then loaded with Ly$\alpha$-emitting gas, and this gas of outflow 
origin lacks the asymmetric morphology characteristic of inflowing streams. 

\cite{Perrotta+19} measure that ERQs show systematically more extreme [\ion{O}{III}]\,$\lambda$5007 
kinematics than blue quasars matched in luminosities and redshifts. Further, among ERQs the 
extreme [\ion{O}{III}]\,$\lambda$5007 kinematics positively correlates with the red $i-W3$ quasar 
colour. In spite of that, in Section 3.5.2 we show that the Ly$\alpha$ halo kinematics of 
J0006+1215 is similar to luminous blue quasars. A lack of dependence of Ly$\alpha$ halo kinematics 
on quasar colour does not bear implications on the dependence between quasar evolutionary stage 
and colour, however. The quiet kinematics indicate that any Ly$\alpha$-emitting materials swept by 
quasar-driven outflows slow to moderate speeds on halo scales in J0006+1215 and in luminous blue 
quasars. The adaptive optics-assisted data in \cite{Vayner+21} show that the extreme 
[\ion{O}{III}]\,$\lambda$5007 outflows of ERQs happen within $\sim$1\,kpc radii. Their result is 
consistent with our wide-field data that show no broadening in Ly$\alpha$ in J0006+1215 down to 
$\sim$0.7\,arcsec or $\sim$6\,kpc radius. Lifetimes of quasar episodes as measured by their 
proximity zones are typically ($10^5$\textendash$10^7$)\,yr \citep[e.g.,][]{Khrykin+21}. Even if 
an observed episode has been on for $10^7$\,yr, for a characteristic outflow speed consistent with 
the kinematics of most Ly$\alpha$ haloes $\sim$400\,km\,s$^{-1}$, an outflow driven by the episode 
can only travel out to $\sim$4\,kpc. Any Ly$\alpha$-emitting materials on circumgalactic scales of 
outflow origin are thus not driven by the observed quasar episode but in the past. The absence of 
fast circumgalactic Ly$\alpha$ outflows in J0006+1215 has no implications on the quasar 
evolutionary stage. 

\subsection{Quasar Illumination Beyond Orientation Effects} 

A fundamental question about the nature of ERQs is whether their obscuration is due to orientation 
effects or a global dust distribution. Orientation effects may be caused by a dusty torus and 
axisymmetric ionization cones as often invoked to explain the Type~I/Type~II dichotomy. On the 
other hand a more global dust distribution may relate to ERQs residing in more dusty and younger 
host galaxies than other luminous quasars. One unusual feature of ERQs is that their spectral 
energy distributions are typically nearly flat in the rest-frame UV despite having extremely red 
colours from rest-frame UV to mid-IR. This spectral shape is not consistent with normal galactic 
reddening curves. It appears to require global patchy dust distributions \citep{Hamann+17}, 
unusually large dust grain sizes as observed in the envelope ejecta of some massive evolved stars 
and supernovae \citep{Hamann12,DiMascia+21}, or a significant contribution from scattered 
rest-frame UV continuum light \citep{Alexandroff+18}. We note that the KCWI data probe 
illumination pattern on circumgalactic scales, not obscuration geometry on circumnuclear scales. 
Our data do not distinguish between ionizing photons escaping through a global patchy dust 
distribution along the line of sight or scattering off an anisotropic dust distribution and then 
escaping along the line of sight. 
In the following we explain that the Ly$\alpha$ halo properties of J0006+1215 suggest unique 
physical conditions for this ERQ that are beyond orientation differences from other known quasar 
populations. 

Figs~\ref{fig:neblum_Lbol} and~\ref{fig:linsz_Lbol} show that the line luminosity and linear size 
of the J0006+1215 \ion{H}{I}\,Ly$\alpha$ halo are well within the ranges spanned by other Type~I 
and Type~II quasar-powered Ly$\alpha$ haloes. The Ly$\alpha$ halo luminosity of J0006+1215 is only 
moderately suppressed compared to similarly intrinsically luminous blue quasars. 
A lack of a trend between halo sizes or halo luminosities and quasar colours can be 
explained if the halo emissions around quasars are not predominantly ionization-bounded 
\citep{DempseyZakamska18}. The amount of halo emission will then only weakly depend on the amount 
of ionizing radiation or obscuration. The only moderately suppresed halo luminosity of J0006+1215 
thus does not imply that J0006+1215 has globally minimal extinction for ionizing photons, as would 
be the case if J0006+1215 were no different from a blue quasar other than orientation. In 
addition, it has been measured that obscured quasars reside in more massive dark matter haloes 
\citep{DiPompeo+17,Geach+19}. Obscured quasars would then have larger amounts of halo gas to 
fluoresce in Ly$\alpha$, which would counteract global extinction effects of ionizing photons. 

Fig.~\ref{fig:HeIILya_CIVLya} shows that the low \ion{He}{II}\,$\lambda$1640/Ly$\alpha$ ratio of 
the J0006+1215 inner halo is closer to those measured for inner haloes around Type~I sources and 
is lower than those of inner haloes around Type~II sources. \cite{Cantalupo+19} discuss how the 
\ion{He}{II}\,$\lambda$1640/Ly$\alpha$ ratio gives clue to the observed illuminated volume of a 
quasar. Considering that the brightness of the high-ionization \ion{He}{II}\,$\lambda$1640 line 
declines with physical distance more steeply than Ly$\alpha$, for a given projected area around a 
quasar, lower \ion{He}{II}\,$\lambda$1640/Ly$\alpha$ ratio implies higher Ly$\alpha$ fluorescent
volume. A quasar episode has finite lifetime, resulting in light travel effects. Under the
Type~I/Type~II dichotomy, the observed Ly$\alpha$ fluorescent volume of a Type~I source which beams
along the line of sight is much larger than that of Type~II source which beams in the plane of the
sky. For a Type~I source, the physical distances spanned by the ionized gas are much larger than
their sky projected distances. The \ion{He}{II}\,$\lambda$1640/Ly$\alpha$ ratio at a given
projected distance from a Type~I source is then lower than that from a Type~II source. It is
intriguing that the J0006+1215 inner halo has a Type~I-like \ion{He}{II}\,$\lambda$1640/Ly$\alpha$
ratio despite our inference that the ionizing radiation is weaker along the line of sight than in
the plane of sky. The low \ion{He}{II}\,$\lambda$1640/Ly$\alpha$ ratio implies the observed
Ly$\alpha$ fluorescent volume along the line of sight is still larger than in the plane of the
sky, which is possible when the fluorescent Ly$\alpha$ emission is not predominantly
ionization-bounded. In turn this implies that the central ionizing photons have certain channels 
to escape along the line of sight. 

Fig.~\ref{fig:eunwght_Lbol} shows that the overall morphology of the J0006+1215 Ly$\alpha$ halo is 
more circularly symmetric than all haloes around Type~II sources. In \cite{denBrok+20} it is 
suggested that under the Type~I/Type~II dichotomy the different orientations of the ionization 
cones imply that Type~II sources should power haloes that appear more asymmetric than Type 
I-powered haloes. The Type~I-like halo morphology of J0006+1215 is consistent with the scenario 
that ionizing photons can escape along the line of sight. 

The Type~I-like halo properties of J0006+1215 are consistent with the appearance of broad emission 
lines in the quasar spectrum clearly shown in Fig.~\ref{fig:KCWIBOSS} despite extremely red 
colour. 

We note that \cite{Goulding+18} find the X-ray absorbing column toward J0006+1215 is in the 
Compton-thin regime, and hence does not suggest anisotropic obscuration.

\subsection{No Evidence of Fast-moving and Ly$\alpha$-emitting Halo Gas}

In Section 3.4 we show that the J0006+1215 inner halo is kinematically quiet with respect to the 
estimated dark matter velocity dispersion. In theory, fast-moving and Ly$\alpha$-emitting halo gas 
are not necessarily expected in radio-quiet quasars at cosmic noon. In the cosmological 
radiation-hydrodynamic simulations of \cite{Costa+22}, the quasar-driven outflowing gas that moves 
at $\sim$1000\,km\,s$^{-1}$ on circumgalactic scales is in a much less dense phase than the bulk 
of the Ly$\alpha$-emitting gas. 
It is very possible that supermassive black hole feedback on the circumgalactic medium becomes 
observationally evident and are required by cosmological hydrodynamic simulations only after 
$z\lesssim2$ \citep{Nelson+19,SoriniDaveAnglesAlcazar20,Sorini+21}. 
An analytical model also proposes that supermassive black hole feedback gently lifts the 
circumgalactic medium and never manifests as fast circumgalactic outflows \citep[e.g.,][]{Voit+20}. 

At present, no selection criteria on extreme properties of radio-quiet quasars systematically find 
fast-moving and Ly$\alpha$-emitting halo gas that are way above the dark matter velocity 
dispersion and require driving by quasars. Although fast Ly$\alpha$ halo gas is occasionally 
detected around single blue quasars, the selection criteria of those quasars do not routinely find 
fast Ly$\alpha$ halo gas. For example, \cite{Ginolfi+18} select a $z\sim5$ quasar by its broad 
absorption lines, and finds high velocity dispersion in the inner Ly$\alpha$ halo. However, the 
broad absorption line quasar in the $z>5.7$ sample of \cite{Farina+19} does not reveal extended 
emission. Another example is that \cite{Travascio+20} select two quasars by their 
hyperluminosities, and finds high velocity dispersion in the inner Ly$\alpha$ halo. However, the 
five hyperluminous quasars among the samples of \cite{Cai+19}, \cite{ArrigoniBattaia+19}, and 
\cite{Borisova+16a} do not display high velocity dispersion in their Ly$\alpha$ haloes. 
It appears that the only selection criterion on central powering source properties that 
systematically finds fast Ly$\alpha$ halo gas is radio loudness, presumably associated with jets 
that extend outside the host galaxies \citep{Swinbank+15,Kolwa+19,Wang+21}. 

\subsection{Future Work} 

The full KCWI sample of ERQs (J.\ Gillette et al.\ in preparation) is needed to sustain the  
findings in this work. We suggest three lines of future work. Firstly, search for fainter, redder 
quasars that are analogues to the dusty first quasars, follow up on their extended emissions, and 
compare with Ly$\alpha$ haloes of the first quasars. Being young and more globally dusty than 
typical $z\sim$\;2\textendash4 blue quasars, together with a characteristic spectral energy 
distribution and thus peculiar extinction properties, it is possible that ERQs are analogues of 
the majority of the first quasars. \cite{Hamann+17} argue for a scenario where ERQs drive 
high-speed outflows that ablate and disperse dusty molecular clouds in their host galaxies, 
resulting in small dusty clumps capable of patchy obscuration on galactic scales. \cite{Ni+22} 
investigate quasar obscuration in a cosmological hydrodynamic simulation, and find that at 
$z\gtrsim6$ much of the obscuration is on galactic scales, i.e.\ does not follow the Type~II 
geometry. \cite{DiMascia+21} investigate obscuration at $z\sim6$ in cosmological hydrodynamic 
simulations, and find a shallow extinction curve. 
\cite{Yung+21} predict UV luminosity functions of active galactic nuclei in a semi-analytic model, 
and finds that dust attenuation effects are increasingly suppressed at higher redshifts. To date, 
Ly$\alpha$ halo surrounding one $z\sim6$ obscured quasar has been detected 
\citep{Connor+19,Farina+19}. Our second suggestion is to search for signatures of quasar feedback 
on circumgalactic scales other than fast flows, especially at later cosmic times. Our third 
suggestion is to search for quasar-driven outflows in line absorption which is more sensitive to 
underdense gas than line emission.

\section{Conclusions}

We present a detailed study of KCWI integral field spectroscopic observations of the reddest ERQ, 
J0006+1215 at $z_{\rm sys}=2.3184$. It has an $i-W3$ colour of 8.01, drives unusally 
fast [{O}{III}]\,$\lambda$5007-emitting outflows on $\sim$1\,kpc scale, and exhibits 
narrow \ion{H}{I}\,Ly$\alpha$ emission in the central 1-arcsec that is otherwise dominated by 
quasar emission. We analyse the central 1-arcsec spectrum and the spatially resolved properties of 
the line-emitting halo. 

We summarize analysis results on the data as follows: 
\begin{itemize}
\item The central 1-arcsec spectrum demonstrates narrow line emissions in Ly$\alpha$, 
\ion{C}{IV}\,$\lambda\lambda$1549,1550, and \ion{He}{II}\,$\lambda$1640 that have similar peak 
velocities and velocity widths. We take the peak of the Ly$\alpha$ spike to be the systemic 
redshift. We detect extended Ly$\alpha$ and \ion{C}{IV}\,$\lambda$1549 and marginally 
spatially resolved \ion{He}{II}\,$\lambda$1640 emissions. 
\item The maximum linear size of detectable Ly$\alpha$ emission is 140\,kpc and that of 
\ion{C}{IV}\,$\lambda$1549 is 47\,kpc. The total Ly$\alpha$ halo luminosity is 
$5.10\times10^{43}$\,erg\,s$^{-1}$. For the Ly$\alpha$ halo the flux-weighted elliptical 
eccentricity with respect to the halo centroid $e_{\rm weight}=0.44$ and the unweighted 
eccentricity with respect to the quasar position $e_{\rm unweight}=0.69$. The halo appears 
circularly symmetric in the inner region and demonstrates some filamentary asymmetry on larger 
scales. The circularly averaged Ly$\alpha$ surface brightness radial profile is centrally 
concentrated, and is best fitted by 
${\rm SB}_{{\rm Ly}\alpha}(r)=(2.46\times10^{14})\times\exp{(-r/(8.5\,{\rm kpc}))}$\,erg\,s$^{-1}$\,cm$^{-2}$\,arcsec$^{-2}$.  
Where both lines in a line ratio are detected, the spatially-integrated ratios 
\ion{C}{IV}\,$\lambda$1549/Ly$\alpha$\,$=0.03$ and 
\ion{He}{II}\,$\lambda$1640/Ly$\alpha$\,$=0.01$. 
\item The kinematics of the extended emissions are quiet, demonstrating no broadening above the 
dark matter circular velocity down to the spatial reolution $\sim$0.7\,arcsec or $\sim$6\,kpc from 
the quasar. On the Ly$\alpha$ map, the spatial median of the velocity centroids of detected 
spaxels is $-82$\,km\,s$^{-1}$, with a spatial standard deviation of 234\,km\,s$^{-1}$. On the 
Ly$\alpha$ map, the spatial median of the velocity dispersions of detected spaxels is 
287\,km\,s$^{-1}$, with a spatial standard deviation of 70\,km\,s$^{-1}$. The peak velocities 
and velocity widths of the spatially-integrated Ly$\alpha$, \ion{C}{IV}\,$\lambda$1549, and 
\ion{He}{II}\,$\lambda$1640 emissions are similar to each other. The inner halo emission is 
kinematically coherent and has blueshifted Ly$\alpha$ absorption, while the outer halo shows 
kinematic shear and has no blueshifted absorption. 
\item The narrow Ly$\alpha$ line profile in a central 1-arcsec aperture and that in the inner halo 
region are very similar. We confirm that the Ly$\alpha$ spike first identified in the BOSS 
spectrum is an inner extension of the halo emission, and is therefore useful as an 
indicator of the systemic redshift. Relative to this halo frame, the broad  
\ion{C}{IV}\,$\lambda$1549 emission in the quasar is blueshifted by 2240\,km\,s$^{-1}$. 
\item The size, line luminosity, and kinematics of the J0006+1215 Ly$\alpha$ halo are within 
ranges spanned by luminous blue quasars at similar redshifts, with the halo luminosity being only 
moderately suppressed relative to blue quasars. The J0006+1215 halo is more centrally 
concentrated and has a more circularly symmetric inner region compared to these other haloes. 
\item The J0006+1215 inner halo has as low \ion{He}{II}\,$\lambda$1640/Ly$\alpha$ ratio as
other Type~I quasars, which is lower than Type~II sources. The J0006+1215 overall halo morphology
is as asymmetric as other Type~I quasars, and less asymmetric than Type~II sources. 
\end{itemize}

We summarize our inferences from the data analysis as follows: 
\begin{itemize}
\item Obscuration in the ERQ allows views of the innermost halo. The narrow Ly$\alpha$ 
component in the central 1\,arcsec and that in the unresolved [\ion{O}{III}]\,$\lambda$5007 
accurately reveal large blueshifts of the broad \ion{C}{IV}\,$\lambda$1549 and broad 
[\ion{O}{III}]\,$\lambda$5007 in the unresolved quasar. 
\item Measured properties of the J0006+1215 halo transition near $\sim$30\,kpc from the centre, 
with the inner region dominated by moderate-speed outflow driven in the past and the outer 
region dominated by inflow. 
\item The quasar illumination pattern and the circumgalactic medium properties of J0006+1215 are 
broadly similar to those of luminous blue quasars. 
\item The higher concentration and the more symmetric appearance of the inner region of the 
J0006+1215 halo are consistent with the ERQ being younger than blue quasars. The 
Ly$\alpha$-emitting outflow materials have not expanded to the outer regions. On the other hand 
the lack of dependence of Ly$\alpha$ halo kinematics on quasar colour despite the strong 
dependence of kpc-scale [{O}{III}]\,$\lambda$5007 kinematics on quasar colour bears no implication 
on evolutionary stage. 
\item As halo emissions are not predominantly ionization-bounded, the only moderately suppressed 
halo luminosity of J0006+1215 does not imply its obscuration is only due to orientation effects. 
The inner halo \ion{He}{II}\,$\lambda$1640/Ly$\alpha$ ratio and the overall halo morphology of  
J0006+1215 being dissimilar to Type~II quasars suggests unique physical conditions for this ERQ 
that are beyond orientation differences from other quasar populations. 
\item While the J0006+1215 ERQ dominates the ionization of the circumgalactic medium, we find no
evidence of mechanical quasar feedback in the Ly$\alpha$-emitting halo. No selection criteria on
any extreme properties of radio quiet quasars have ever systematically found fast-moving 
Ly$\alpha$ halo gas that requires driving by the quasars.          
\end{itemize}

\section*{Acknowledgements}

We are grateful to our collaborator A.\ Vayner. MWL thanks KCWI instrument team members M.\ 
Matuszewski and J.\ D.\ Neill for technical support. MWL thanks A.\ Battaia, Z.\ Cai, 
S.\ Cantalupo, Y.\ Chen, Y.\ T.\ Chow, E.\ Farina, Y.\ Guo, E.\ T.\ Hamden, Q.\ Li, Z.\ Li, and 
D.\ B.\ O'Sullivan for helpful discussions. MWL, FH, and JG acknowledge support from the USA 
National Science Foundation grant AST-1911066. NLZ is supported at the Institute for Advanced 
Study by the J.\ Robert Oppenheimer Visiting Professorship and the Bershadsky Fund, and 
acknowledges support from NASA ADAP grant 80NSSC21K1569. This research was supported in part by 
the National Science Foundation under Grant No.\ NSF PHY-1748958. 
This work is based on observations made at the W.\ M.\ Keck Observatory, which is operated as a 
scientific partnership between the California Institute of Technology and the University of 
California. It was made possible by the generous support of the W.\ M.\ Keck Foundation. Data 
presented herein were partially obtained using the California Institute of Technology Remote 
Observing Facility. The authors wish to recognize and acknowledge the very significant cultural 
role and reverence that the summit of Mauna Kea has always had within the indigenous Hawaiian 
community. We are most fortunate to have the opportunity to conduct observations from this 
mountain. 

\section*{Data Availability}
 
The data are available upon request. 

\bibliographystyle{mnras}
\bibliography{/Users/lwymarie/Documents/Bibli/allrefs}


\bsp	
\label{lastpage}
\end{document}